%% file: ms.tex
\documentclass[11pt,a4paper]{article}
\pdfoutput=1

\usepackage{amsmath, amssymb}
\usepackage[T1]{fontenc}
\usepackage{jheppub}
\usepackage{psfrag}
\usepackage{slashed}
\usepackage{cancel}
\usepackage{lscape}
\usepackage{soul}
\usepackage{physics}
\usepackage{hyperref}
\usepackage{subcaption}
\usepackage{array}
\usepackage{graphicx}
\usepackage{subcaption}
\usepackage{multirow}
\usepackage{float}
\usepackage{makecell}
\usepackage{tikz}
\usetikzlibrary{automata, positioning}
\usepackage{colortbl}
\usepackage[nameinlink,capitalize,noabbrev]{cleveref}
\usepackage[]{siunitx}
\usepackage[export]{adjustbox}
\usepackage[utf8]{inputenc}

\newcommand\njet{\texttt{NJet}}
\newcommand\fastjet{\texttt{FastJet}}
\newcommand\herwig{\texttt{Herwig}}
\newcommand\pythia{\texttt{Pythia}}
\newcommand\sherpa{\texttt{Sherpa}}
\newcommand\python{\texttt{Python}}
\newcommand\matplotlib{\texttt{Matplotlib}}
\newcommand\numpy{\texttt{Numpy}}
\newcommand\pandas{\texttt{Pandas}}
\newcommand\cpp{\texttt{C++}}
\newcommand\eigen{\texttt{Eigen}}
\newcommand\rivet{\texttt{Rivet}}
\newcommand\madgraph{\texttt{MadGraph}}
\renewcommand\to{\rightarrow}

\def\incite#1{Ref.~\cite{#1}}

\definecolor{mygreen}{rgb}{0,0.7,0}

\preprint{{IPPP/20/116}}

\title{Optimising simulations for diphoton production at hadron colliders using amplitude neural networks}

\author[a,b]{Joseph Aylett-Bullock}
\author[c]{Simon Badger}
\author[a]{Ryan Moodie}

\affiliation[a]{
Institute for Particle Physics Phenomenology, Department of Physics, Durham University, Durham, DH1
3LE, United Kingdom%
}
\affiliation[b]{
Institute for Data Science, Durham University, Durham, DH1
3LE, United Kingdom%
}
\affiliation[c]{
Dipartimento di Fisica and Arnold-Regge Center, Universit\`{a} di Torino,
and INFN, Sezione di Torino, Via P. Giuria 1, I-10125 Torino, Italy
}
\emailAdd{j.p.bullock@durham.ac.uk, simondavid.badger@unito.it, ryan.i.moodie@durham.ac.uk}

\abstract{Machine learning technology has the potential to dramatically optimise event
generation and simulations. We continue to investigate the use of neural networks to approximate matrix elements
for high-multiplicity scattering processes. We focus on the case of loop-induced diphoton production through gluon fusion,
and develop a realistic simulation method that can be applied to hadron collider observables. Neural
networks are trained using the one-loop amplitudes implemented in the \njet~\cpp~library,
and interfaced to the \sherpa~Monte Carlo event generator, where we perform a detailed study for
$2\to3$ and $2\to4$ scattering problems. We also consider how the trained networks perform when varying the
kinematic cuts effecting the phase space and the reliability of the neural network simulations.
}


\begin{document}
\maketitle
\flushbottom

\input{introduction}
\input{diphotonamps}
\input{setup}
\input{results}
\input{conclusions}

\newpage
\input{appendix}

\newpage
\section*{Acknowledgements \label{sec:acknowledgements}}

We thank Frank Krauss for many insightful discussions, Alan Price and Marek Schoenherr for
assistance with \sherpa~and Joseph Walker for help with \rivet.  This project received funding from
the European Union's Horizon 2020 research and innovation programmes \textit{High precision
multi-jet dynamics at the LHC} (grant agreement No 772009). JAB is supported by the UK Research and
Innovation Science and Technology Facilities Council (UKRI-STFC) grant number ST/P006744/1 and ST/P001246/1. RM is
supported by UKRI-STFC ST/S505365/1 and ST/P001246/1. This paper made use of \python~\cite{python}
and the following \python~libraries: \matplotlib~\cite{matplotlib}, \numpy~\cite{numpy} and
\pandas~\cite{pandas1,pandas2}.

\bibliographystyle{JHEP}
\bibliography{bib}

\end{document}

%% file: introduction.tex
\section{Introduction \label{sec:introduction}}

Phenomenological studies of high multiplicity final states at collider
experiments present a substantial theoretical challenge and are increasingly
important ingredients in experimental measurements. During the last 15 years, a
dramatic improvement in computational algorithms for one-loop amplitudes has
led to a number of highly automated codes capable of predictions at
next-to-leading order (NLO) accuracy in the Standard Model (SM)~\cite{Berger:2008sj,Bevilacqua:2011xh,Cullen:2014yla,Alwall:2014hca,Denner:2017wsf}.

These codes are based around numerical algorithms which bypass the growth in algebraic
complexity that analytic approaches suffer from. The computational cost of
these algorithms is however relatively high, resulting in huge commitments of
CPU and personnel resources to obtain the necessary theoretical predictions for
current experiments.

The interface of general one-loop amplitude codes into multi-purpose
Monte Carlo (MC) event generators has resulted in a wide variety of simulation
options which can offer the best possible theoretical accuracy. Methods that go
beyond fixed-order perturbation theory --- such as parton shower matching, 
merging, and jet multiplicities --- improve accuracy across important regions of
phase space. However, these simulations add additional strain on the underlying amplitudes.

State-of-the-art tools make use of advanced phase-space mapping algorithms to
improve the convergence of the multi-dimensional integration. General purpose
MC event generators such as \sherpa~\cite{Gleisberg:2008ta, Bothmann:2019yzt}, \pythia~\cite{Sjostrand:2006za, Sjostrand:2014zea}, \herwig~\cite{Bahr:2008pv, Corcella:2000bw, Bellm:2015jjp},
and \madgraph~\cite{Alwall:2014hca} often
make use of the diagram structure of the underlying tree-level process to
ensure an optimal distribution of points.
Reusing tree-level distributions when
generating virtual events is particularly effective at reducing the
computational cost of using complicated one-loop amplitudes.

In this paper, we will consider a class of scattering processes that contribute
to diphoton signals at hadron colliders. The process $gg\to \gamma\gamma+n(g)$
is a good test case for machine learning (ML) technology, since it is loop
induced and has relevant contribution from high multiplicity matrix elements. Using automated tools at NLO, full QCD corrections are known for
$pp\to\gamma\gamma\,\,+\leq3$ jets
\cite{Gehrmann:2013bga,Badger:2013ava,Bern:2014vza}. There has been a flurry
of recent activity around next-to-next-to-leading-order (NNLO) corrections to $pp\to\gamma\gamma+j$ in which
the complete leading colour corrections have been
presented~\cite{Agarwal:2021grm,Chawdhry:2021mkw,Agarwal:2021vdh,Chawdhry:2021hkp,Badger:2021imn}.
As such, $2\to 3$ and $2\to 4$ scattering for the gluon-initiated diphoton
channel are now extremely relevant for future phenomenological studies.

ML is now extremely popular in high energy physics with a wealth
of applications. The majority of problems concern classification and exploit
the scalability of ML algorithms to large datasets. The fact that neural networks (NNs)
offer a general function parameterisation that are also useful in regression
problems is not new in particle physics either: for example, parton distribution functions (PDFs) produced by the NNPDF collaboration~\cite{Ball:2014uwa} have been used for many years by the LHC experiments.

Interpolation methods such as polynomial fits and interpolation grids have
often been used to provide fast simulations and predictions of differential
observables, particularly at NNLO where there is an intricate structure of infrared (IR) singularities \cite{Czakon:2008zk, Borowka:2016ehy, Heinrich:2017kxx,
Jones:2018hbb, Heinrich:2019bkc}. These techniques are extremely powerful for problems with two or three hard scattering variables,
but scale poorly at high multiplicity.

ML offers a possible solution to the poor scaling with larger number of
variables. Various directions are currently being explored for different
aspects of event generation and MC integration. 
Boosted decision trees (BDTs) and NNs have been applied to phase-space sampling and integration \cite{Bendavid:2017zhk, Klimek:2018mza, Chen:2020nfb} and have shown to increase the speed at which functions can be integrated, as well as allow for the integration of functions for which traditional algorithms failed. 
More recently, sampling using normalising
flows~\cite{rezende2015variational, Dinh2014NICENI} has been attempted \cite{Gao:2020vdv, Bothmann:2020ywa, Gao:2020zvv}. These have been shown to avoid the computational cost of calculating the gradient of the network itself, when determining the Jacobian, which previous NN approaches required.

There has also been a large focus on using ML for other components of MC event generator simulations. Specifically, Generative Adversarial Networks (GANs) \cite{NIPS2014_5423} are being applied to event generation \cite{Otten:2019hhl, Hashemi:2019fkn, DiSipio:2019imz, Butter:2019cae, Carrazza:2019cnt, SHiP:2019gcl, Butter:2020tvl, Butter:2020qhk, Alanazi:2020jod, Butter:2020abv, Alanazi:2020klf, Lebese:2021foi}, event unweighting \cite{Backes:2020vka, Verheyen:2020bjw} and subtraction \cite{Butter:2019eyo}, with recent works incorporating Bayesian methods for uncertainty estimation into these generative methods \cite{Bellagente:2021yyh}. NN-based approaches (some of which also use GAN technology) applied to parton showering \cite{Bothmann:2018trh, deOliveira:2017pjk, Monk:2018zsb, Dohi:2020eda} and event reweighting \cite{Nachman:2020fff} have also been developed.

Several works have focused on developing NN techniques for explicitly learning the cross section of a given process \cite{Otten:2018kum, Buckley:2020bxg}; however, little research has been done on learning the matrix element itself for a given phase-space point and process. \incite{Bishara:2019iwh} took a parallelised BDT approach to learning the matrix element, and tested this on the loop-induced $gg\to ZZ$ process at leading order, demonstrating the potential for large speedups in matrix element calculations. An ensembled NN methodology was presented in \incite{Badger:2020uow} which divides the phase space into divergent and non-divergent regions, and further splits the former into sub-regions corresponding to different IR singular structures. This approach was tested on $e^+e^-\to\,\,\leq\,5\, \text{jets}$ at both tree and one-loop level, to ensure robustness at high multiplicity, and was found to provide a good approximation of the cross section and various differential distributions.

Readers may like to refer to the living review of ML in particle physics for up-to-date information about the state-of-the-art \cite{Feickert:2021ajf}.

Our paper is organised as follows. We begin by giving a brief review of the structure of the underlying perturbative amplitudes that we study. We then describe the computational setup in which we use the \njet~amplitude library~\cite{Badger:2012pg} to train an ensemble of NNs that are then interfaced to the \sherpa~MC event generator~\cite{Gleisberg:2008ta, Bothmann:2019yzt}. We discuss a strategy for the generation of the training dataset, and a reweighting strategy to refine the distributions of events where the networks are used for matrix element generation. We then discuss results for $gg\to\gamma\gamma g$ and $gg\to\gamma\gamma gg$ at leading order (LO). In each case, we present a selection of differential observables and provide a detailed comparison with a conventional simulation. We end with a presentation of our conclusions and a discussion of possible future uses of these technologies.

Our code is publicly available at \url{https://github.com/JosephPB/n3jet_diphoton}.

%% file: diphotonamps.tex
\section{Gluon-initiated diphoton amplitudes}

We study amplitudes with two photons and many gluons which first appear at one-loop level in the SM.
With conventional simulations relying on cheaper LO tree evaluations to optimise event generation for NLO one-loop contributions, these loop-induced processes present an interesting sector to test new approaches for phase-space integration.
Compact analytic computations for $gg\to \gamma\gamma$ and $gg\to \gamma\gamma g$ have been available for some time and offer extremely fast and stable evaluation.
As a result, it is feasible to optimise event generation with the one-loop evaluation.
For $2\to 4$ scattering problems, only numerical codes are available and simulations can be extremely slow.
It also not clear that analytic formula would be sufficiently compact to alleviate this situation even if they were available.
For this reason, we consider an alternative setup where the whole simulation uses a NN approximation of the amplitude.

The loop-level amplitudes proceed through a fermion loop and have a colour decomposition in the trace basis as
\begin{multline}
  \mathcal{A}^{(1)}(1,\ldots,n-2,(n-1)_\gamma,n_\gamma) = \\
  {g_s}^{n-2} g_{\gamma\gamma} \sum_{\sigma\in S_{n-3}} \lambda(\sigma\left(a_1, \ldots, a_{n-2}\right)) A^{(1)}(\sigma(1,\ldots,n-2),(n-1)_\gamma,n_\gamma),
\end{multline}
where $g_s$ is the strong coupling, $g_{\gamma\gamma} = e^2\sum_q Q_q^2$ is the combined coupling of the diphoton system to the fermion loop, $S_{n-3}$ is the set of even non-cyclic permutations of $\{1,\ldots,n-2\}$, $t^a$ are the fundamental $\mathrm{SU(3)}$ generators, $q$ runs over active quark flavours with fractional quark charge $Q_q$, and the colour trace function $\lambda$ is defined as
\begin{equation}
  \lambda\left(a_1, \ldots, a_{n-2} \right) = {\rm tr}\left( t^{a_1}t^{a_2}\ldots t^{a_{n-2}} \right) + (-1)^n {\rm tr}\left( t^{a_1}t^{a_{n-2}}\ldots t^{a_{2}} \right).
\end{equation}
This yields $(n-3)!/2$ primitive amplitudes $A$ for $n\geq5$.
For example, for $n=4$ there is a single primitive amplitude.
It is given by the diagrams
\begin{equation}
A(1,2,3_\gamma,4_\gamma) =
\includegraphics[valign=c]{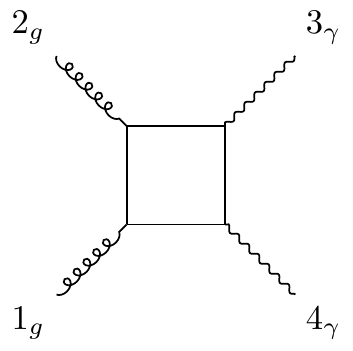}
+
\includegraphics[valign=c]{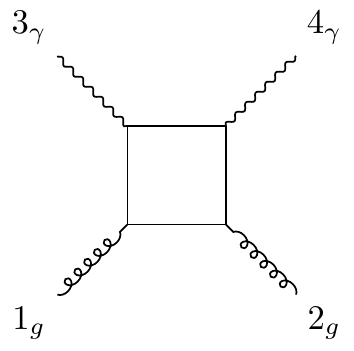}
+
\includegraphics[valign=c]{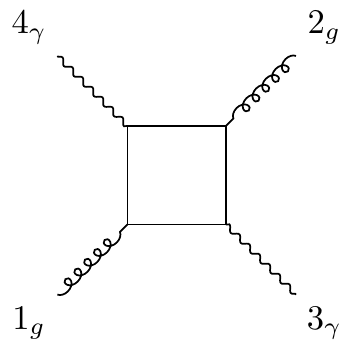}
\end{equation}
\noindent where a plain line indicates a sum over quark loop arrow directions. At one-loop, these amplitudes are also related to the fermion loop corrections to pure gluon scattering through permutations \cite{deFlorian:1999tp}. The ingredients for differential cross sections are the squared amplitudes summed over helicities, $h$, and colour,
\begin{equation}
  {\lvert\mathcal{A}^{(1)}\rvert}^2 = \left(\frac{\alpha_s}{4\pi}\right)^{n-2} g_{\gamma\gamma}^2 \sum_{h,i,j} {A_i^{(1)}}^{*}(h)\, \mathcal{C}_{ij}\, A_j^{(1)}(h) + \mathcal{O}(\alpha_s^{n-1})
  \label{eq:colhelsum}
\end{equation}
where the matrix $\mathcal{C}$ is a function of the number of colours, $N_c$, obtained by squaring the colour basis elements and the index on the partial amplitudes, $A$, refers to the different permutations in the colour decomposition.

The amplitudes in this article are taken from the \njet~\cpp~library \cite{Badger:2012pg}. Here, there are different options: a general numerical setup using generalised unitarity and integrand reduction; and hard-coded analytic expressions for $n=4,5$. 
The $n=4$ analytic expressions were taken from \incite{Bern:2001df}, while for $n=5$ they were obtained directly from a finite field reconstruction \cite{Peraro:2019svx} and are in agreement with known analytic formula \cite{deFlorian:1999tp,Bern:1993mq}. By using a momentum twistor parameterisation of the external kinematics, cancellations in the rational coefficients of the special functions that lead to a manifestly finite representation are easily identified.

The numerical evaluation requires the sum of permutations of ordered primitive amplitudes.
This is completely automated for arbitrary multiplicity, but evaluation times and numerical stability are increasingly difficult to control.

\begin{figure}[htp]
    \centering
    \includegraphics[width=0.6\textwidth]{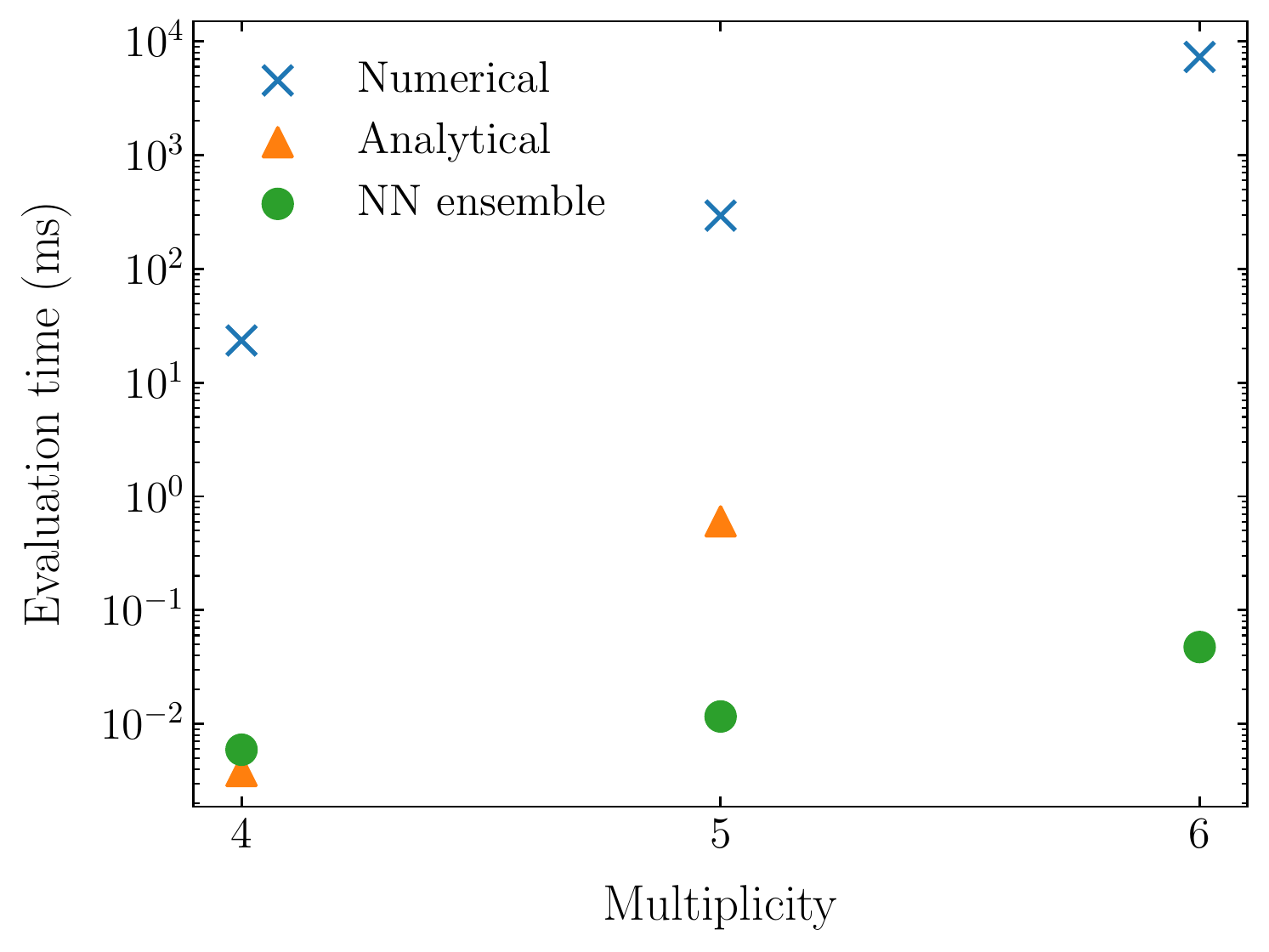}
    \caption{
    Matrix element typical CPU evaluation times for available methods --- including \njet~numerical evaluations, \njet~analytical evaluations, and inference on a NN ensemble as described in \cref{sec:setup} --- against the number of legs.
    These calls are single-threaded as parallelisation is applied at the level of events in simulations.
    An analytic expression for $2\to4$ is not available.
    The NN is comparable to the analytic call at $2\to2$, 50 times faster at $2\to3$, then $10^5$ times faster than the $2\to4$ numeric call.
    }
    \label{fig:timing-basic}
\end{figure}

To study the growth of evaluation time with multiplicity, we evaluate the matrix element at 100 random phase-space points with each available technique and plot the mean times in \cref{fig:timing-basic}.
We generate the phase-space points isotropically with the algorithm from \incite{Byckling:1971vca}.
While analytic methods are competitive at low multiplicity, we see they scale poorly and are unlikely to beat numerics at $n\ge6$.
Numeric scaling is better, but these algorithms come with a high cost.
Our NN approach provides a performant alternative, with significantly better scaling than either numerics or analytics.

%% file: setup.tex
\section{Computational setup}
\label{sec:setup}

In this paper, we build on previous work which sought to demonstrate the viability of using NN-based approaches to approximate matrix element values for hard scattering processes \cite{Badger:2020uow}. In that work, a NN ensemble approach was presented in which a different NN is trained on each soft and collinear region of phase space, and was shown to be effective in handling IR divergent structures at both the Born and one-loop level at high multiplicity in $e^+e^-$ collisions. We extend this to more complex $2\rightarrow 3$ and $2\rightarrow 4$ gluon-initiated diphoton amplitudes, while also showing the ability for these ML models to interface with existing event generators such as \sherpa~\cite{Gleisberg:2008ta, Bothmann:2019yzt}. This is important to demonstrate since it is not immediately obvious that NN approximations trained in isolation will be robust to the added intricacies of event generators which are important for extracting physical results, such as PDF weighting and choices of integrators.

We begin this section with an overview of the setup for training our ML approach based on those presented in \incite{Badger:2020uow} and finish with a discussion on interfacing with event generators. 

\subsection{Phase-space partitioning}
\label{sec:partitioning}

IR divergences arise from soft and collinear real emissions and integrals over massless partons appearing in virtual corrections. Our ML approach requires the isolation of real emission IR singularities such that a different NN can be trained on each soft and collinear region \cite{Badger:2020uow}. To do this, we partition the phase space into divergent and non-divergent regions and then subsequently sub-divide the divergent region according to the FKS subtraction \cite{Frixione:1995ms,Frederix:2009yq}. The original implementation of this method was for $e^+e^- \rightarrow q\bar{q} + \text{jets}$ collisions in which the IR singularities only appear in the final state. In the processes discussed in this paper, IR singularities appear in initial-initial, initial-final and final-final state pair combinations. We therefore extend the partitioning to these states for hadronic processes as specified in \incite{Frederix:2009yq}.

As in \incite{Badger:2020uow}, we parameterise our phase space according to the Lorentz invariant $y_{ij} = s_{ij}/s_{12}$, where $s_{ij} = (p_i + p_j)^2$.
The PDF convolution creates a non-fixed partonic centre-of-mass energy $\sqrt{s_{12}}$.

We now define the divergent and non-divergent regions of phase space as

\begin{align}
\mathcal{R}_{\text{div}} &= \left\{p\,|\, \text{min}(y_{ij}) \leq  y_{p} ,\, p = (p_1,p_2,\ldots,p_n),\,i,j \in \{1,\ldots,n\}\right\}, \label{eqn:R_div} \\
\mathcal{R}_{\text{non-div}} &= \left\{p\,|\, y_{p} \leq \text{min}(y_{ij}) ,\, p = (p_1,p_2,\ldots,p_n),\,i,j \in \{1,\ldots,n\}\right\}, \label{eqn:R_non_div}
\end{align}
 
\noindent where $y_{p}$ is fixed for the process (see \cref{app:delta_tuning}), $p$ is a phase-space point consisting of the incoming momenta $\{p_1,p_2\}$, and the outgoing momenta $\{p_3,\ldots,p_n\}$, where $n$ is the number of particles in the process. The FKS pairs are then defined as

\begin{multline}\label{FKS pairs}
\mathcal{P}_{\text{FKS}} = \{(i,j)\,\,|\,\,1 \leq i \leq n,\,\,2 \leq j \leq n,\,\, i \neq j, \\
\mathcal{M}^{(n,0)}\,\, \text{or} \,\,\mathcal{M}^{(n,1)}\to\infty\,\, \text{if} \,\,p_i^0\to 0\,\, \text{or} \,\,p_j^0\to 0\,\, \text{or} \,\,\vec{p}_i\parallel\vec{p}_j \}.
\end{multline}

Finally, partition functions are used as in \incite{Badger:2020uow}

\begin{equation}\label{S_i,j definition}
\mathcal{S}_{i,j} = \frac{1}{D_{1}s_{ij}}, \,\,\,\,\,D_1 = \sum_{i,j \in \mathcal{P}_{\text{FKS}}} \frac{1}{s_{ij}},
\end{equation}

\noindent such that

\begin{equation}\label{FKS_cs_add}
\text{d}\sigma = \sum_{i,j} S_{i,j}\, \text{d}\sigma,
\end{equation}

\noindent where $\sigma$ represents the cross section for the process. These partition functions are then used to weight the corresponding matrix elements of points lying in $\mathcal{R}_{\text{div}}$, and we train one network in this region for each $S_{i,j}$ (see \cref{sec:nn_setup} and \incite{Badger:2020uow}).

The above definitions are appropriate for the processes we studied here as they account for the different singularity structure than that found in the case of $e^+e^-$ (i.e. they explicitly include the initial-state singularities). To allow for easier generalisability to other processes, we include all pairs of initial- and final-state particles in our implementation, including the $\{\gamma\gamma\}$ pair which is redundant as it does not exhibit the relevant singularity structure. This redundancy does increase the computational time required; however, we find the performance of the NN ensemble is not adversely affected. The above implementation could therefore be simply generalised to the $e^+e^-$ case, although again with some redundancy.

\subsection{Neural network setup}
\label{sec:nn_setup}

Now that we have described how we partition the phase space for data generation, we will discuss the NN setup, and the data processing required to train and test our methodology, in more detail. While the focus of this paper is the use of the NN ensemble method originally presented in \incite{Badger:2020uow}, for completeness we present a comparison of this method against a naive (single network) approach in \cref{app:single_comparison}.

\subsubsection{Data}
\label{sec:data}

The sampling of phase space is dependent on the integrator.
Unless otherwise specified, the same integrator is used for training, validation and testing.
We generate the datasets from two runs of the integrator:
the first is divided into training and validation datasets according to a 80:20 split;
the second uses a different random seed than the first and is used for the test dataset (note that this second stage is only performed when evaluating the performance of our ML approach).
Phase-space points, and their corresponding matrix elements generated by \njet, are extracted during these stages after the cuts have been applied.
For consistency with \incite{Badger:2020uow}, we train on 100k points and test on 3M.
Input data consists of the initial-state 4-momenta, $x \in \mathbb{R}^{4n}$, and the target data is the corresponding (weighted) matrix element, $y \in \mathbb{R}$.

Once the training data has been generated, we split the data into divergent and non-divergent regions according to \cref{eqn:R_div} and \cref{eqn:R_non_div}, and create copies of the points in $\mathcal{R_{\text{div}}}$, each weighted by $\mathcal{S}_{i,j}$ for $(i,j)\in\mathcal{P}_{\text{FKS}}$, which are used to train a different network in the NN ensemble for each soft and collinear divergent structure.

Before training the NN ensemble, we preprocess the data by standardising each variable input, i.e.~we ensure that the training and validation input distributions for each variable have a mean of zero and a standard deviation of one. The output variable distribution is similarly standardised. During the testing phase, we insure the testing data is standardised to the training and validation dataset distribution parameters as is customary in ML literature. Once the model has been used for matrix element estimation, the output is destandardised to obtain the final value. Varying types of data processing are used for hyperparameter tuning (see \cref{app:hyperparameter} for more details).

All our simulations use $\sqrt{s_{12}}=1\,\text{TeV}$; the methodology is agnostic to this choice. Although we test the performance of our models on different cuts, unless otherwise specified all models are trained, and analyses performed, using the following kinematic cuts adapted from those in \incite{Badger:2013ava}

\begin{align*}
p_{T,j} &> \SI{20}{GeV} & R_{\gamma,j} &> 0.4 & \abs{\eta_j} &< 5 \\
p_{T,\gamma_1} &> \SI{40}{GeV} & R_{\gamma,\gamma} &> 0.4 & \abs{\eta_{\gamma}} &< 2.37 \\
p_{T,\gamma_2} &> \SI{30}{GeV} & & & & \\ 
\end{align*}

\noindent where $p_T=\sqrt{{p_x}^2+{p_y}^2}$ (beam along $z$-axis) is transverse momentum magnitude, $R=\sqrt{(\Delta\eta)^2+(\Delta\phi)^2}$ is isolation cut cone radius, $\eta$ is pseudorapidity, $\phi$ is azimuthal angle, $\gamma$ denotes a photon, photons are ordered by $p_T$, and jets, $j$, are identified through the anti-$k_T$ algorithm \cite{Cacciari:2008gp} implemented in \fastjet~\cite{Cacciari:2011ma} with $R=0.4$.
These cuts are typical for LHC analyses.
Photons are selected by smooth cone isolation \cite{Frixione:1998jh} such that all cones of radius $r_\gamma<R$ satisfy
\begin{align*}
    E_{\text{hadronic}}(r_\gamma)\leq\epsilon\,p_{T,\gamma}\frac{1-\cos{r_\gamma}}{1-\cos{R}}
\end{align*}
\noindent with $R=0.4$ and $\epsilon=0.05$.

Matrix elements are evaluated with renormalisation scale $\mu_R = m_Z$ with physical constant values $\alpha$($Q^2=0$), $\alpha_s(m_Z)$, and $m_Z$ from the PDG \cite{10.1093/ptep/ptaa104}.
Since the one-loop process is LO, the full amplitude is finite and has $\mu_R$ dependence in the couplings only. 

\subsubsection{Architecture}

For optimal results, a different NN architecture construction would be fine-tuned to each choice of setup parameters, e.g.~integrator, cuts, and process.
However, the required computational and time resources required to perform this optimisation make this highly impractical.
Instead, we further test the generalisability of the hyperparameter choices made in \incite{Badger:2020uow} to this new set of processes.
In addition, hyperparameter tuning was performed to assess how optimal this approach is relative to the ideal scenario where hyperparameters are process specific, and found the original setup to be among the most optimal (see \cref{app:hyperparameter} for more details).

In summary, we use the same fully-connected NN architecture for every network in the ensemble.
These are parameterised using Keras \cite{keras} and a TensorFlow \cite{tensorflow} backend, with the number of input nodes equal to $n\times4$.\footnote{Testing was performed to assess the change in performance when removing redundant, non-independent, 4-momenta components; however, this had little effect.}
The hidden layers comprise of 20-40-20 nodes and there is a single output node.
All hidden layers use hyperbolic-tangent activation functions and the output node has a linear activation function.
A mean squared error loss function is used, and the network is optimised using Adam optimisation \cite{adam}.
Finally, the number of training epochs is determined through Early Stopping (see Section 8.1.2 of \incite{goodfellow}), tracking the validation loss with no minimum change requirements.
As in \incite{Badger:2020uow}, to minimise the limiting effects of using a validation set containing only 20\% of the original training set, we train with a patience of 100 epochs. 

\subsection{Interfacing with event generators}

Assessing performance after interfacing with existing event generator technology is important for demonstrating `real-world deployment' of ML algorithms in particle physics simulations, as it exposes the model to a range of post-inference effects which may alter the final reliability of the model. For example, generators allow for the easy implementation of complex phase-space cuts, jet clustering algorithms, phase-space and PDF weights, as well as different integrators and integration optimisation routines. 

\subsubsection{The interface}

Event generators are largely written in \cpp~for computational efficiency. Therefore, after the model has been trained, the weights of each NN are extracted and written to file. A \cpp~program reads these models files and performs the linear algebra operations required during the inference step using \eigen~\cite{eigenweb}. This means the \python~libraries used for model training are circumvented, and the call time for model inference is reduced, while keeping everything in \cpp~simplifies the interfacing of the model with standard event generators.

 Given a set of 4-momenta, a custom \cpp~interface provides the helicity- and colour-summed matrix element to \sherpa. This can be used to call \njet~evaluations through a BLHA interface \cite{Binoth:2010xt,Alioli:2013nda} or to call the model inference result. \rivet~\cite{Buckley:2010ar,Bierlich:2019rhm} is then used for analysis, using a script adapted from the reference analysis of \incite{Aaboud:2017vol}.

\subsubsection{Phase-space integration}
\label{sec:integration}

Phase-space integrators seek to achieve increasingly optimal rates of integration convergence though the careful sampling of points. While the choice of integrator can affect the overall rate of convergence, it also determines the placement of phase-space points which directly feeds into the distribution of points in the training dataset. 
 
Since these processes are loop-induced, for simplicity we use the RAMBO integrator \cite{Kleiss:1985gy} throughout for event generation. However, we test different approaches to generating the integration grid. The first we term the `unit grid' which is constructed by running the grid optimisation step while returning a unit value in place of the matrix elements. This effectively removes the dependence on the optimisation procedure and, since RAMBO is used, ensures a uniformly and isotropically sampled phase space. The second uses VEGAS \cite{Lepage:1980dq, Ohl:1998jn} optimisation when generating the integration grid, thereby putting a preference on sampling regions of particular importance to the cross section. We share the integration grid between training and testing phases, meaning this importance sampling is reflected in both. Given the expense of matrix element calculation for the $2\rightarrow4$ scattering process, we reserve the use of VEGAS optimisation only for the $2\rightarrow3$ case.

\subsubsection{Weights}

When training the models, we do not include explicitly any event generator effects.
All additional weightings, i.e.~phase-space weights and PDF weights, are introduced after the model has been used for inference as is done for other matrix element generators.
The addition of these weightings has the potential to be problematic for model performance: when the model is trained it is unlikely to learn all regions of phase space equally well and there is a chance that those regions in which the model has poor performance could be amplified by these additional weighing factors.

In order to test for this we include PDF weights using the LHAPDF library \cite{Buckley:2014ana} and the NNPDF3.1 set {\verb NNPDF31_nlo_as_0118 } \cite{Ball:2017nwa} as well as phase-space weights which depend on the integration grid optimisation method.

\subsection{Reweighting}\label{sec:reweighting}

The approach used in this paper to train the NN ensemble provides good agreement between the network output and that of \njet.
However, the ensemble approach will always be an approximation and is subject to perform poorly in certain regions of phase space, especially those in which it has not been trained on or in which training data is sparse.
As a partial remedy to this, we propose the idea of reweighting the event weights with known matrix element values derived from \njet.
When using weighted event generation this can either be performed after event generation or can be done `on the fly' at the interface level.
This former approach is possible since the original event weight can be recovered through

\begin{equation}
    w_{E, i}^{(\njet)} = w_{E, i}^{(\text{NN})} \times \frac{\mathcal{M}_{i}^{(\njet)}}{\mathcal{M}_{i}^{(\text{NN})}},
\end{equation}

\noindent where $w_{E,i}^{\njet}$ and $w_{E,i}^{\text{NN}}$ are the event weights using \njet~and the NN ensemble respectively for a given phase-space point $i$, and $\mathcal{M}_i^{\njet}$ and $\mathcal{M}_i^{\text{NN}}$ are the associated matrix elements.

As the ratio $\njet/\text{NN}$ is not known a priori, we must construct criteria on which to reweight. Specifically, we explore the following: 

\begin{enumerate}
    \item A random sample of points (e.g. 10\%) regardless of where they are in phase space;
    \item A priori stating which regions of phase space in which to reweight and then doing so either randomly or over the entire region;
    \item Using the NN uncertainties to inform reweighting, e.g.~points with large uncertainties are reweighted.
\end{enumerate}

There are several factors informing which approach is the most appropriate. The first of these is the added compute time required: all of these techniques necessitate the calculation of the matrix element by an analytic or numerical evaluator and therefore limit the desirable number of points to reweight.
The second is the performance gain and confidence in the output in certain regions of phase space.
If the analysis being performed is specific to an under-sampled region of phase space, such as distribution tails where the network may under-perform due to divergent structures in the matrix element, this could be an especially important region in which to reweight.
However, if general process explorations are being performed, meaning all distributions and cross sections are of relative equal importance, then a less restrictive reweighting on regions of phase space may be optimal.

In this paper, we explore the application of reweighting to the processes described in \cref{sec:results}.
While reweighting is not always found to be necessary given the performance of our methodology, we demonstrate how it can be applied and discuss which reweighting criteria show the greatest performance gain.

%% file: results.tex
\section{Results}\label{sec:results}

In this section, we present the results of our experiments for the $2\rightarrow3$ and $2\rightarrow4$ gluon-initiated diphoton amplitudes. As the former is significantly less computationally expensive, we use this for a deep analysis and exploration. The proposed pipeline for using our ML set up and interface with event generators is as follows:

\begin{enumerate}
    \item Generate an integration grid;
    \item Use this with a matrix element provider to generate training and validation datasets;
    \item Train the model;
    \item Use the model to estimate the values of the remaining phase-space points for event generation while using the same integration grid;
    \item Reweight as necessary;
    \item Obtain final results.
\end{enumerate}

\noindent To assess performance, we also evaluated matrix elements with \njet~in parallel with the models, with different random seeds.

\subsection{\texorpdfstring{$gg \rightarrow \gamma \gamma g$}{3g2A}}\label{sec:results_5}

\begin{figure}[h]
    \centering
    \includegraphics[width=0.5\textwidth]{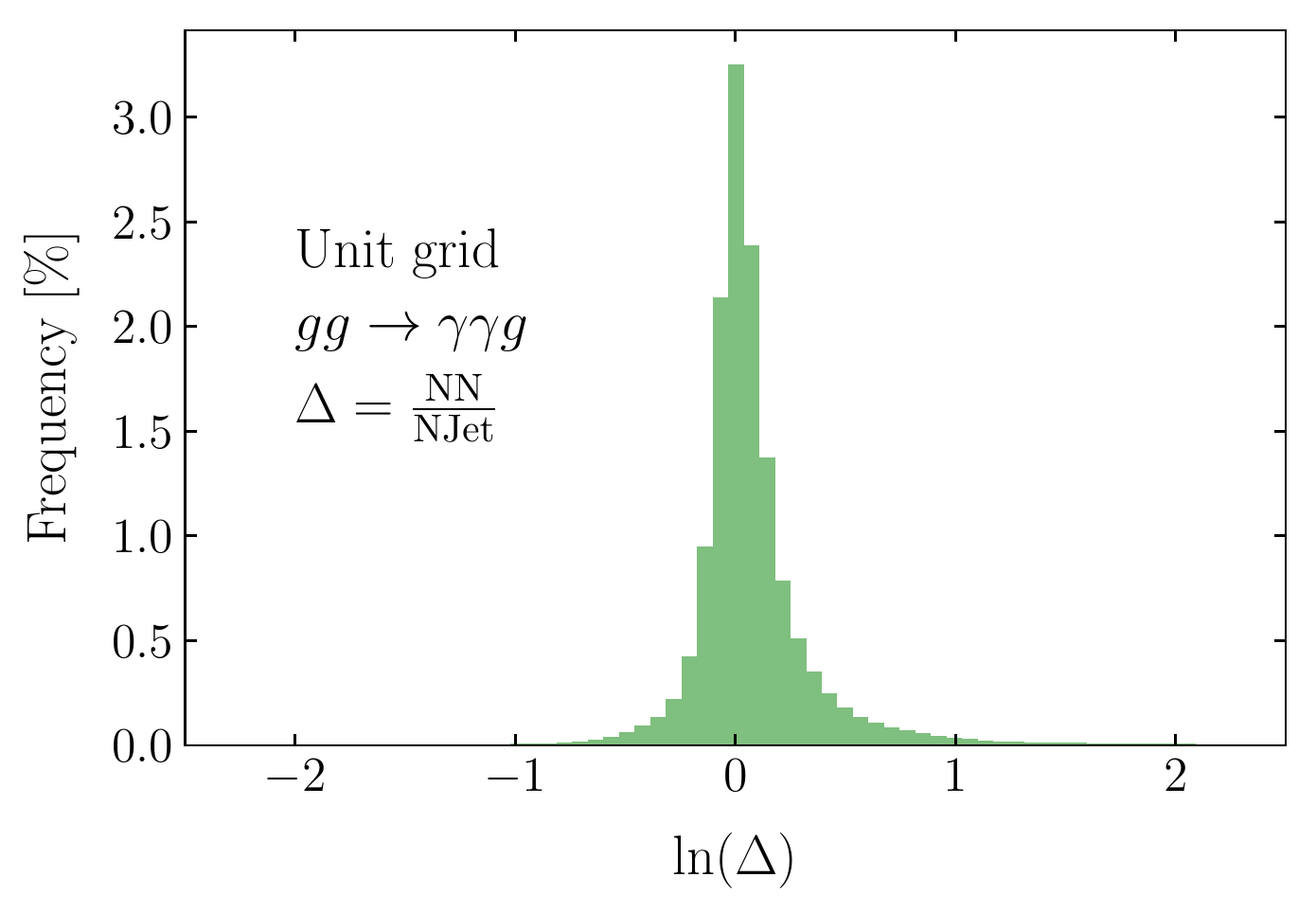}
    \caption{NN/\njet~errors for the $2\rightarrow3$ scattering process using a unit integration grid.}
    \label{fig:5_error_unit}
\end{figure}

First we investigate the performance of our methodology on the loop-induced $gg \rightarrow \gamma \gamma g$ process. Following the procedure outlined above, we use a unit integration grid, choose a random seed to generate the training and validation datasets with, and use another to infer on the trained model.

\cref{fig:5_error_unit} shows the performance of our trained NN ensemble at the matrix element level, here represented as the ratio of the model inferred values to the \njet~evaluations.
The errors form a narrow and approximately symmetric unit-centered distribution, thus demonstrating that the ensemble method has a reasonable per-point accuracy.
The slightly elongated right tale of the distribution is due to large matrix element values in highly divergent regions of phase space, yet these points are in the minority.

\begin{table}[t]
\centering
    \begin{tabular}{ |c|c|c| }
    \hline
    Cuts & \njet~[pb] & NN ensemble [pb] \\
    \hline
    Baseline & $4.149 \times 10^{-6} \pm 6 \times 10^{-9}$ & $4.19 \times 10^{-6} \pm 7 \times 10^{-8}$ \\
    Baseline + $p_{T,\gamma} > 50\,\text{GeV}$ & $5.283 \times 10^{-7} \pm 8 \times 10^{-10}$ & $5.4 \times 10^{-7} \pm 2 \times 10^{-8}$ \\
    Baseline + $m_{\gamma,\gamma} > 50\,\text{GeV}$ & $3.300 \times 10^{-6} \pm 5 \times 10^{-9}$ & $3.34 \times 10^{-6} \pm 5 \times 10^{-8}$ \\
    \hline
    \end{tabular}
    \caption{Cross-sectional comparison between \njet~and the NN ensemble approach using different cuts. Baseline cuts are those specified at the beginning of \cref{sec:results}. The \njet~results are quoted with MC errors and the NN ensemble results with precision/optimality uncertainties calculated as described in \incite{Badger:2020uow}.}
    \label{tab:5_XS}
\end{table}

Once the ensemble is trained, it is converted to be called by the event generator interface which allows for the calculation of the cross section and differential distributions.
\cref{tab:5_XS} shows the results of the cross section derived using \njet~and the NN ensemble.
We see that these two approaches are in excellent agreement, with the ensemble result overlapping within one standard deviation of that calculated by \njet.
The errors on the \njet~values are the MC errors, and the errors on the ensemble are precision/optimality uncertainties.
The latter are calculated by training multiple ensembles with different random seeds in the weight initialisation, and in the shuffling of the training and validation datasets.
MC errors are quoted to one standard deviation and the precision/optimality uncertainties to one standard error on the mean.
A more in depth description of this uncertainty analysis can be found in Section 2.3 of \incite{Badger:2020uow}.

The error plot and cross-section calculation provide good evidence for the performance of the NN ensemble method both in its ability to learn the distribution of phase-space points on average, as well as its robustness to being integrated into a wider event generation framework with additional phase-space and PDF weights.
To further test the methodology in a more relevant way to how it would be used in practice, differential distributions can be used to assess robustness as they more explicitly expose performance on the divergent and tail events.

\begin{figure}[H]
     \centering
     \begin{subfigure}[b]{0.42\textwidth}
         \centering
         \includegraphics[width=\textwidth]{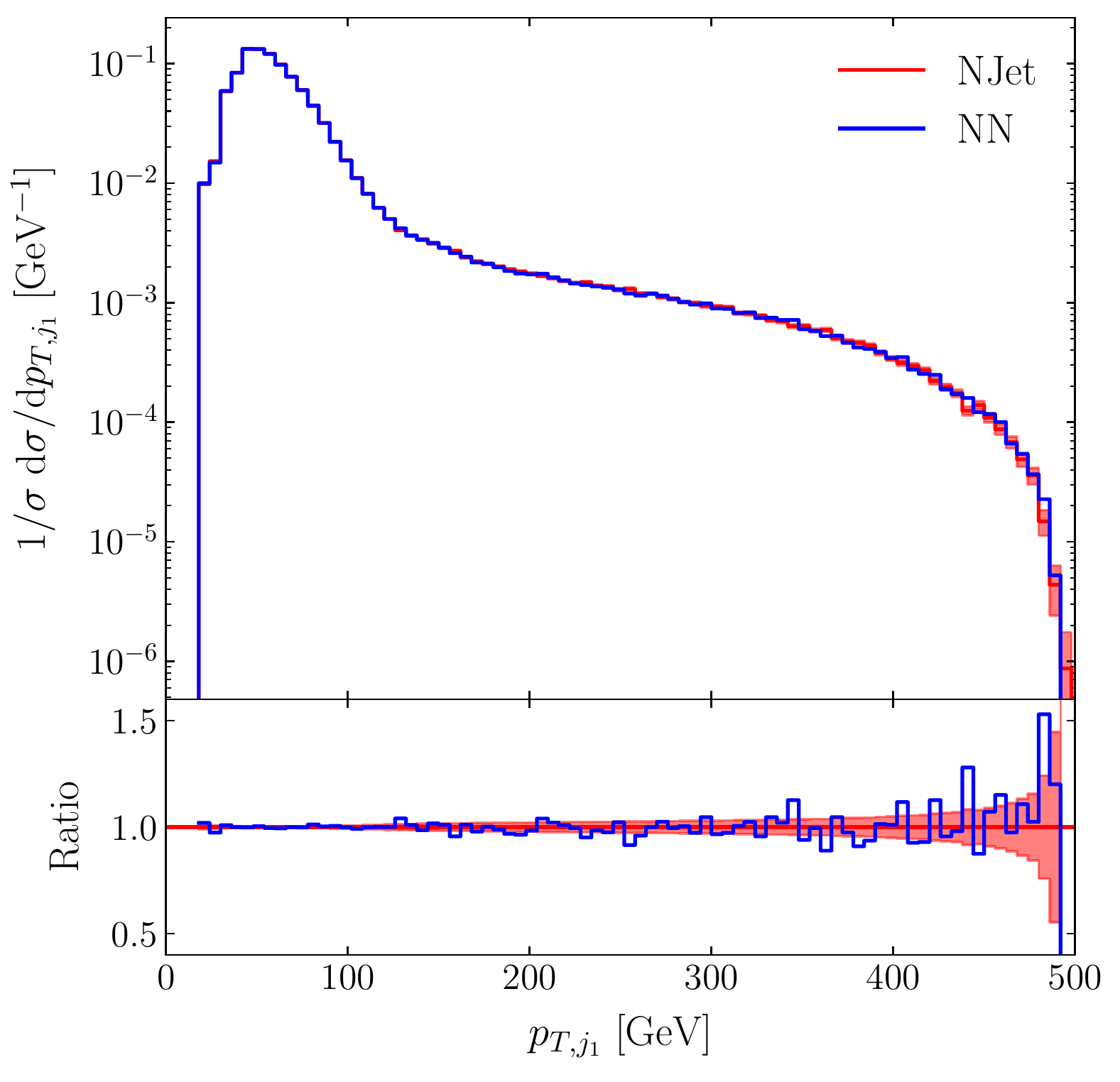}
     \end{subfigure}
     \hfill
     \begin{subfigure}[b]{0.42\textwidth}
         \centering
         \includegraphics[width=\textwidth]{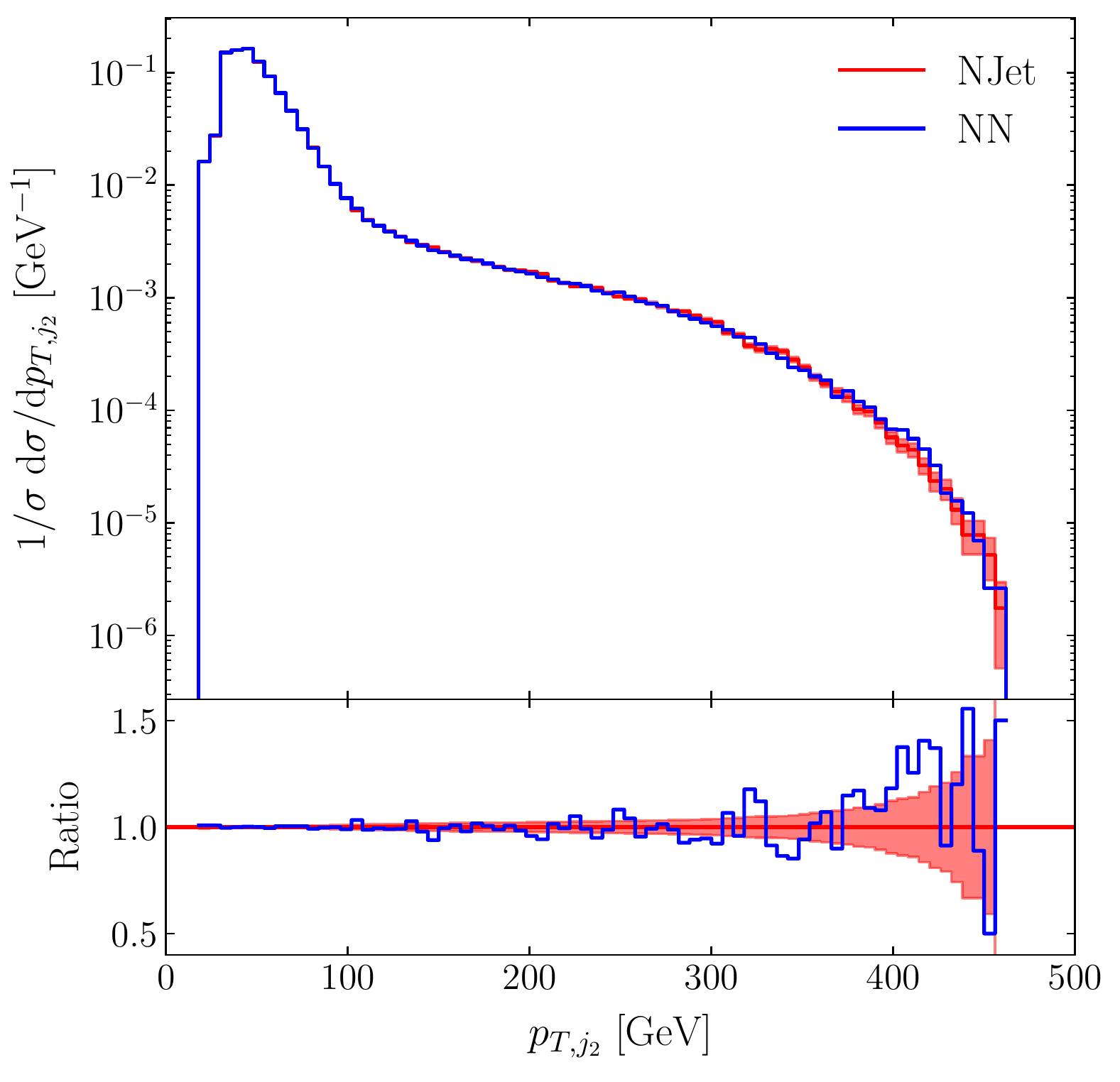}
     \end{subfigure}
     \vfill
     \begin{subfigure}[b]{0.42\textwidth}
         \centering
         \includegraphics[width=\textwidth]{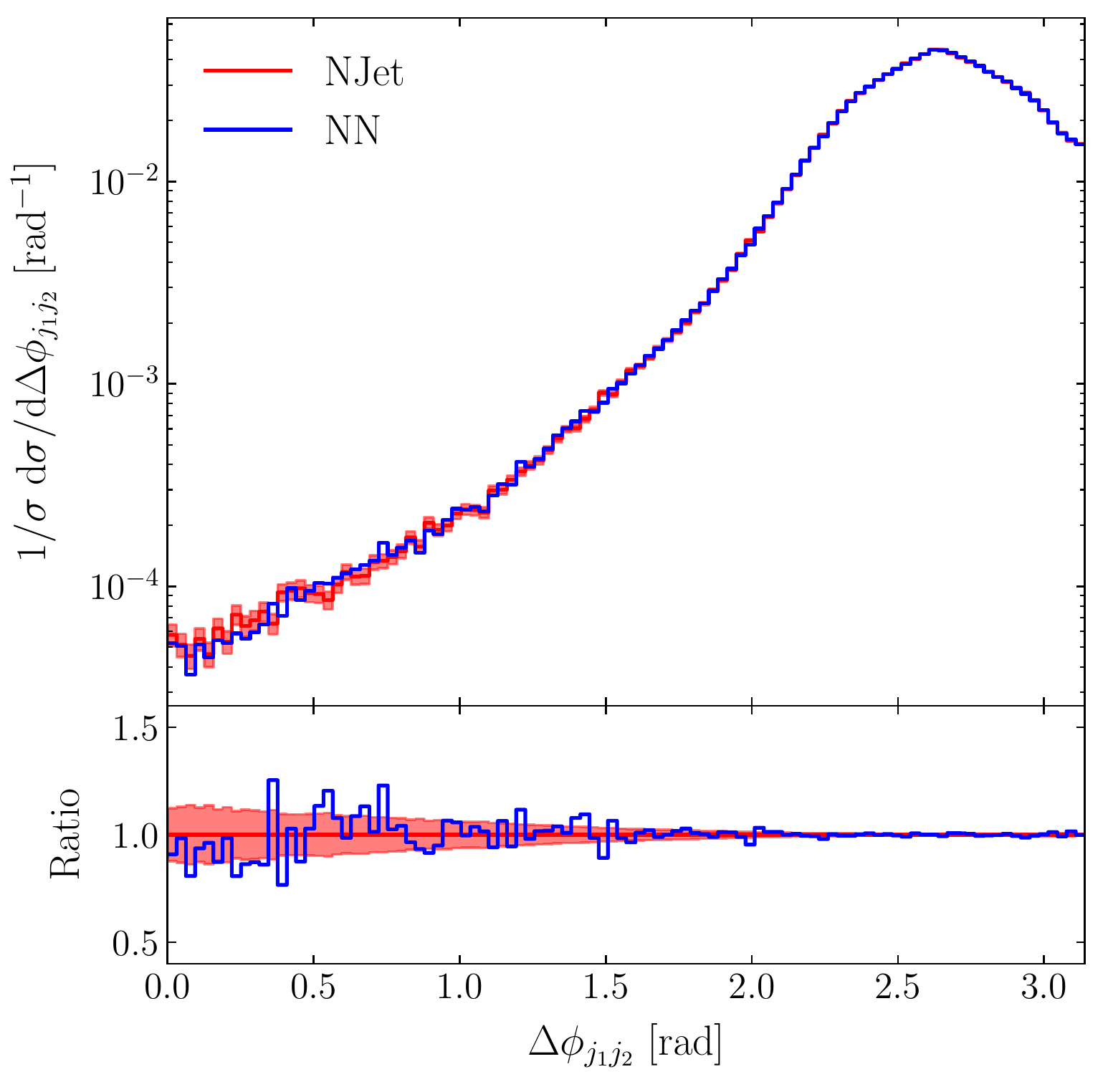}
     \end{subfigure}
     \hfill
     \begin{subfigure}[b]{0.42\textwidth}
         \centering
         \includegraphics[width=\textwidth]{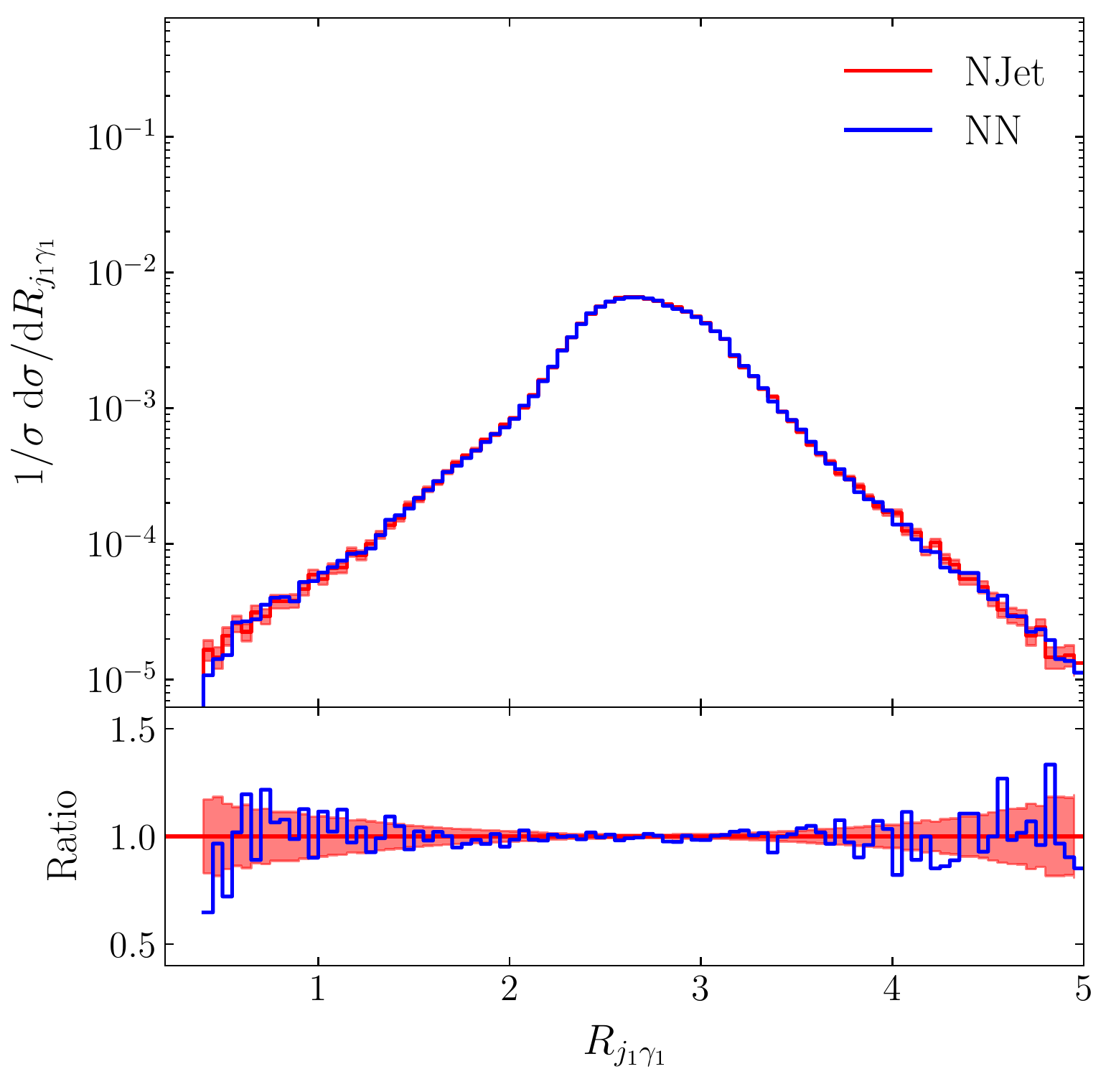}
     \end{subfigure}
     \vfill
     \begin{subfigure}[b]{0.42\textwidth}
         \centering
         \includegraphics[width=\textwidth]{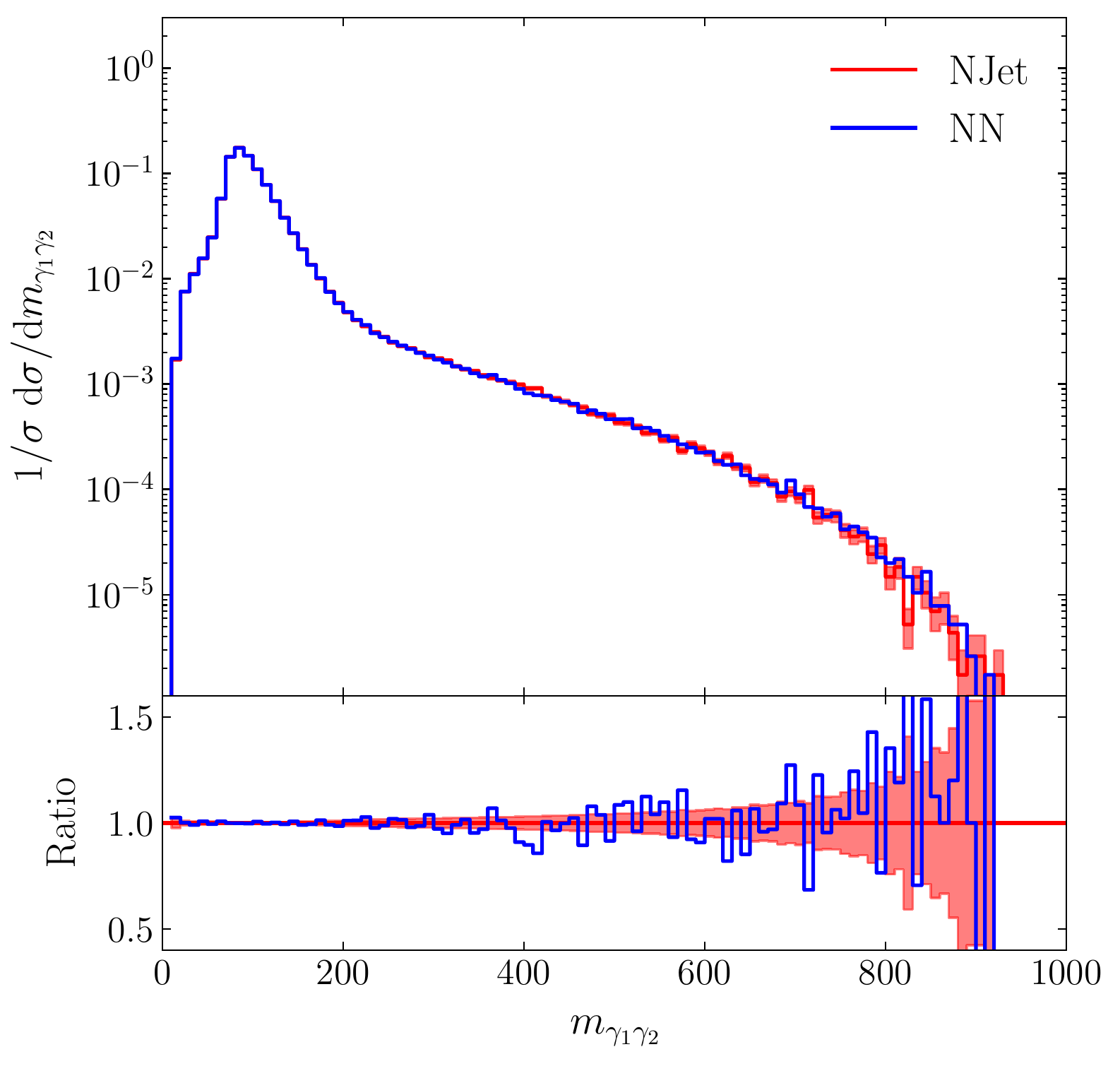}
     \end{subfigure}
     \hfill
     \begin{subfigure}[b]{0.42\textwidth}
         \centering
         \includegraphics[width=\textwidth]{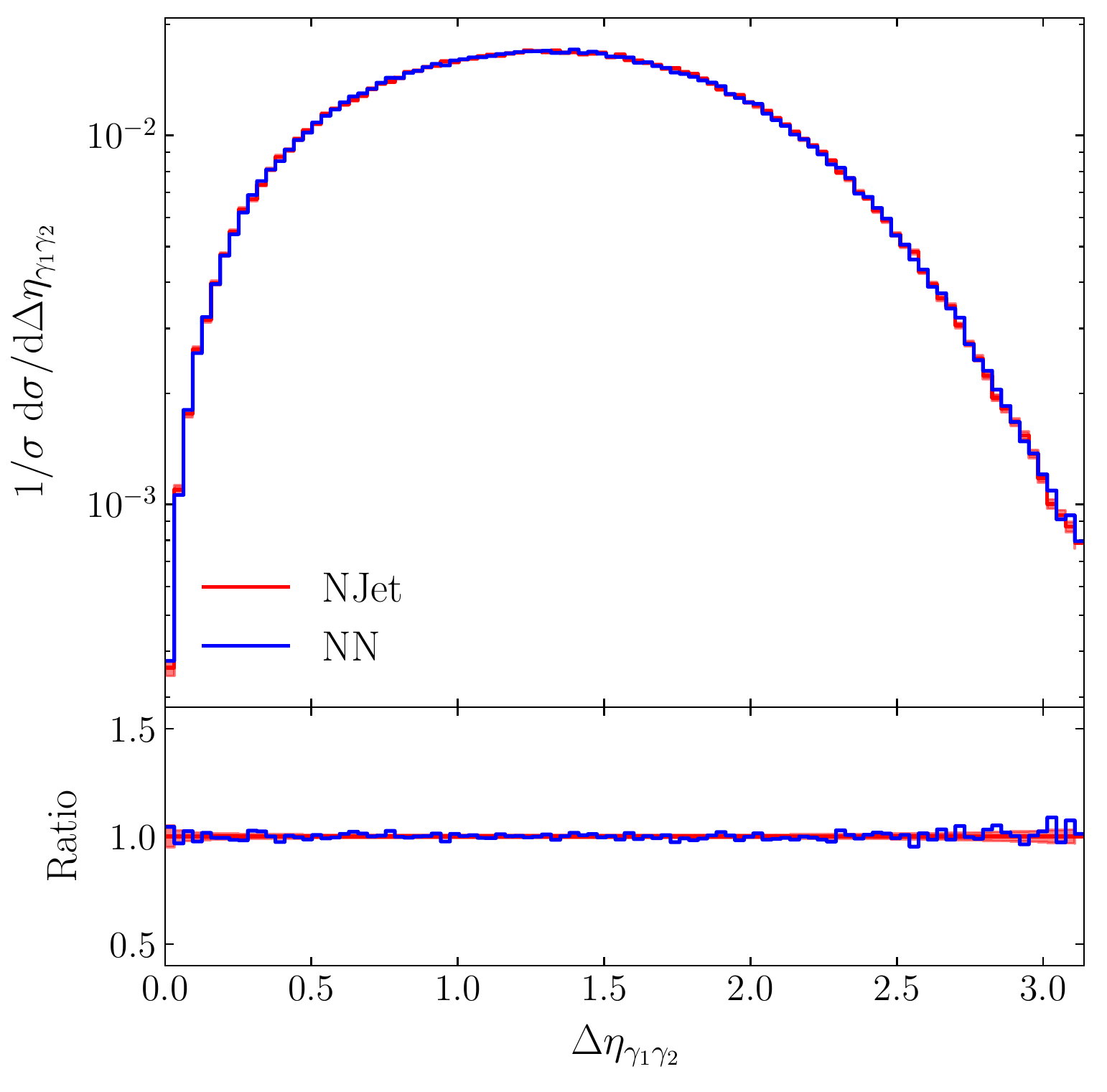}
     \end{subfigure}
     \caption{
     Differential distributions normalised to the cross section for the $2\rightarrow3$ process comparing \njet~(red) with the NN ensemble (blue).
     The \njet~results are quoted with MC errors, and the NN results with precision/optimality uncertainties calculated as described in \incite{Badger:2020uow} but which are negligible in comparison.
     Pseudojets $j_i$ and photons $\gamma_i$ are ordered by energy, $\Delta\phi$ is azimuthal separation, $R$-separation is defined in \cref{sec:data}, and $m_{\gamma_1,\gamma_2}$  and $\Delta\eta_{\gamma_1,\gamma_2}$ are the mass and pseudorapidity separation of the diphoton system.
     }
     \label{fig:5_differentials}
\end{figure}

\cref{fig:5_differentials} demonstrates the performance of the NN ensemble in comparison to \njet~in six differential slices of phase space. These include $p_T$, angular, and diphoton system distributions which have been chosen to give a range of realistic constructions exploring different regions of phase space. In general, the NN ensemble is found to be in good agreement, particularly around the peaks, with the majority of the NN bin values being with the \njet~MC error. The normalised NN uncertainties on the differential bins is negligible in comparison to the MC error. Strong performance is pronounced in the pseudorapidity distribution which shows variation at the percent level. The $p_T$ and angular distributions show more fluctuations in the tail events, with the diphoton mass demonstrating the greatest deviations in these regions. However, despite these differences, fluctuations are clearly statistical rather than systematic meaning agreement will increase as the bins are aggregated. This is to be expected given the strong cross-section performance.

The results presented so far have been derived from a NN ensemble trained and tested on the same integration grid and on the same cut parameters. However, in phenomenological explorations it is common to study a range of cut parameters, especially when measuring the effects of new phenomena. Since the NN ensemble performs well at the per-point level (as shown in \cref{fig:5_error_unit}), it should also be able to generalise to different cut parameter configurations. Specifically, the ensemble should still be applicable to harsher cuts than those used in training because the it expects the training and testing datasets to be drawn from the same statistical distributions. However, in the event that cuts are relaxed in comparison to those the model was trained on, reweighting could be employed for the relevant additional subset of points thereby guaranteeing the expected values in these `unseen' regions of phase space.

\begin{figure}
    \centering
    \includegraphics[width=0.6\textwidth]{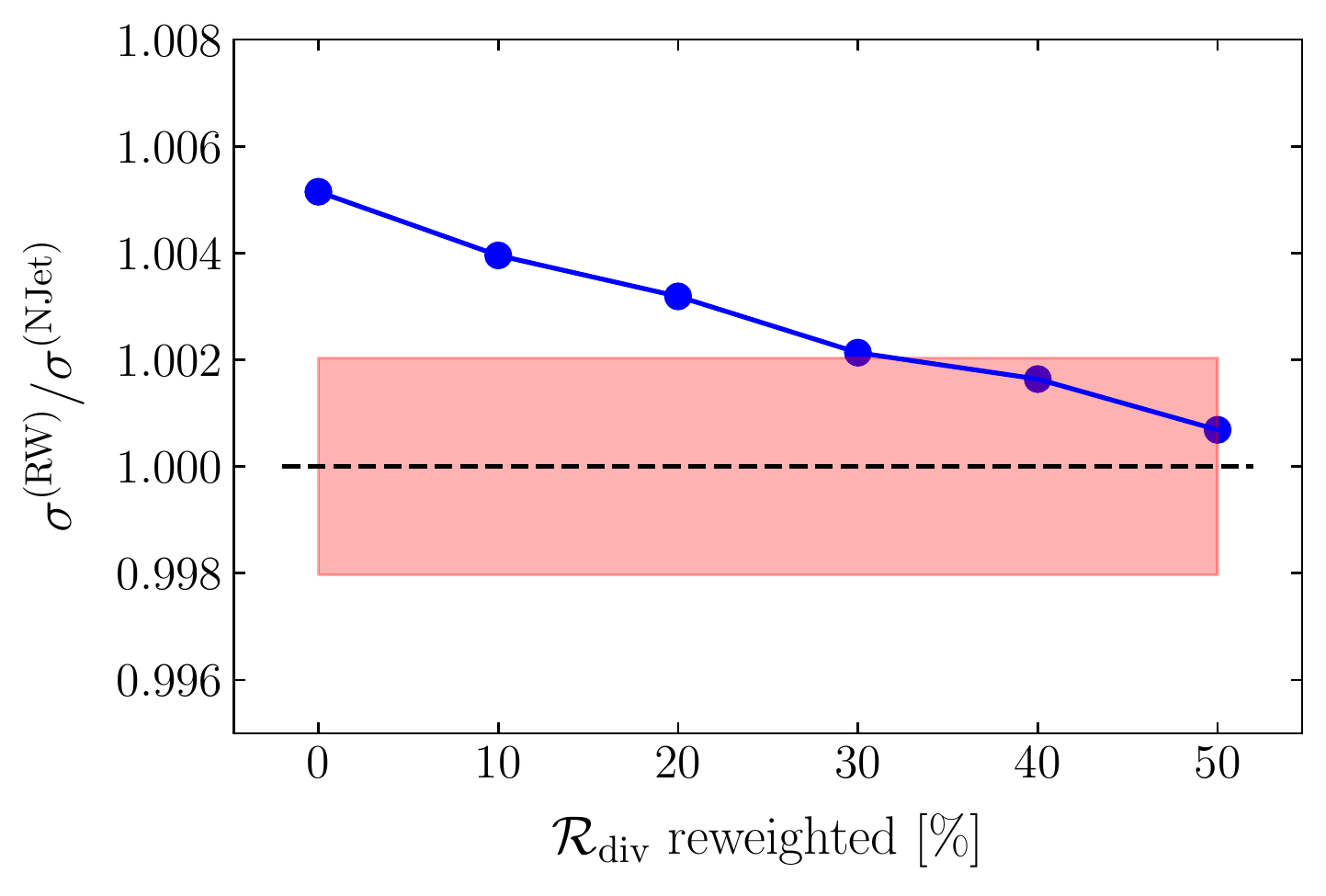}
    \caption{Effect of reweighting points in the divergent region of phase space, $\mathcal{R}_{\text{div}}$, on the ratio between the reweighted cross section, $\sigma^{\text{(RW)}}$, and the cross section calculated using \njet~$\sigma^{\text{(\njet)}}$ for the $2\rightarrow3$ process. In this case, the divergent region comprises approximately 7--9\% of the total phase space (see \cref{app:delta_tuning} for details). The red band shows the MC error on the \njet~result.}
    \label{fig:5_reweighting}
\end{figure}

\cref{tab:5_XS} presents a comparison of cross-section values calculated using \njet~and the NN ensemble with harsher cut values than the baseline. The agreement between the two approaches is comparable to the agreement found before the additional cuts were added, thereby suggesting good generalisability. Indeed, this is not surprising since the points with the largest errors between the NN and \njet~were the most divergent points and therefore the ones more likely to be cut given the IR singularities present in these processes.

The generalisation to additional cut parameters both demonstrates the robustness of this training regime, as well as the practical gain in not having to retrain a network for each specified set of cuts. This allows us to generalise the training and testing procedure outlined at the beginning of this section to suggest that the NN ensemble be first trained on more relaxed cuts and then, as iterations of harsher cut parameters are explored during analysis, these can be applied without the ensemble significantly decreasing in performance. If cuts are to be relaxed then reweighting could be used to ensure good performance at the expense of compute time.

While the network performance has been shown to be strong overall, other reweighing methods can still be explored. Reweighting randomly across all phase space, even at the 20--40\% level, was not found to significantly reduce the difference in the computed cross sections. Similarly, the NN ensemble uncertainties were not found to be correlated with the errors, and so were discarded as a good reweighting criteria. As mentioned above, the points in which targeted reweighting can be most beneficial are those which fall within the divergent regions of phase space. \cref{fig:5_reweighting} presents the results of reweighting points randomly in $\mathcal{R}_{\text{div}}$ (as defined in \cref{eqn:R_div}), and shows an improvement in the cross section --- reweighting a greater number of points enables the reweighted cross section, $\sigma^{\text{(RW)}}$, to converge to the value calculated by \njet, $\sigma^{\text{(\njet)}}$. Indeed, to achieve almost equal values in the cross sections, the total proportion of phase space requiring reweighting is at the percent level. Therefore, we find reweighting in the $\mathcal{R}_{\text{div}}$ region of phase space and/or when relaxing cuts in relation to those used during training can improve model performance.

\begin{figure}
     \centering
     \begin{subfigure}[b]{0.49\textwidth}
         \centering
         \includegraphics[width=\textwidth]{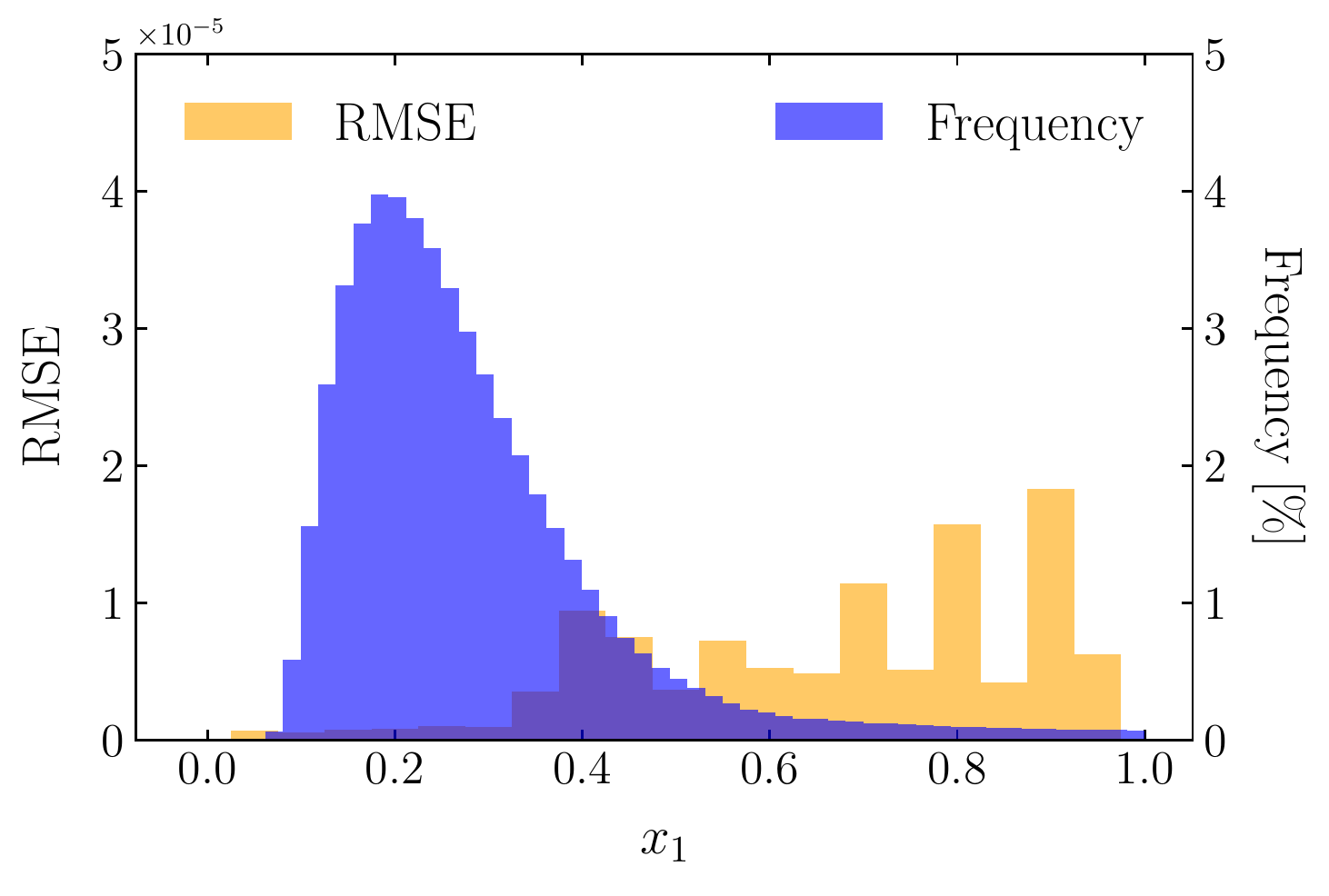}
     \end{subfigure}
     \hfill
     \begin{subfigure}[b]{0.49\textwidth}
         \centering
         \includegraphics[width=\textwidth]{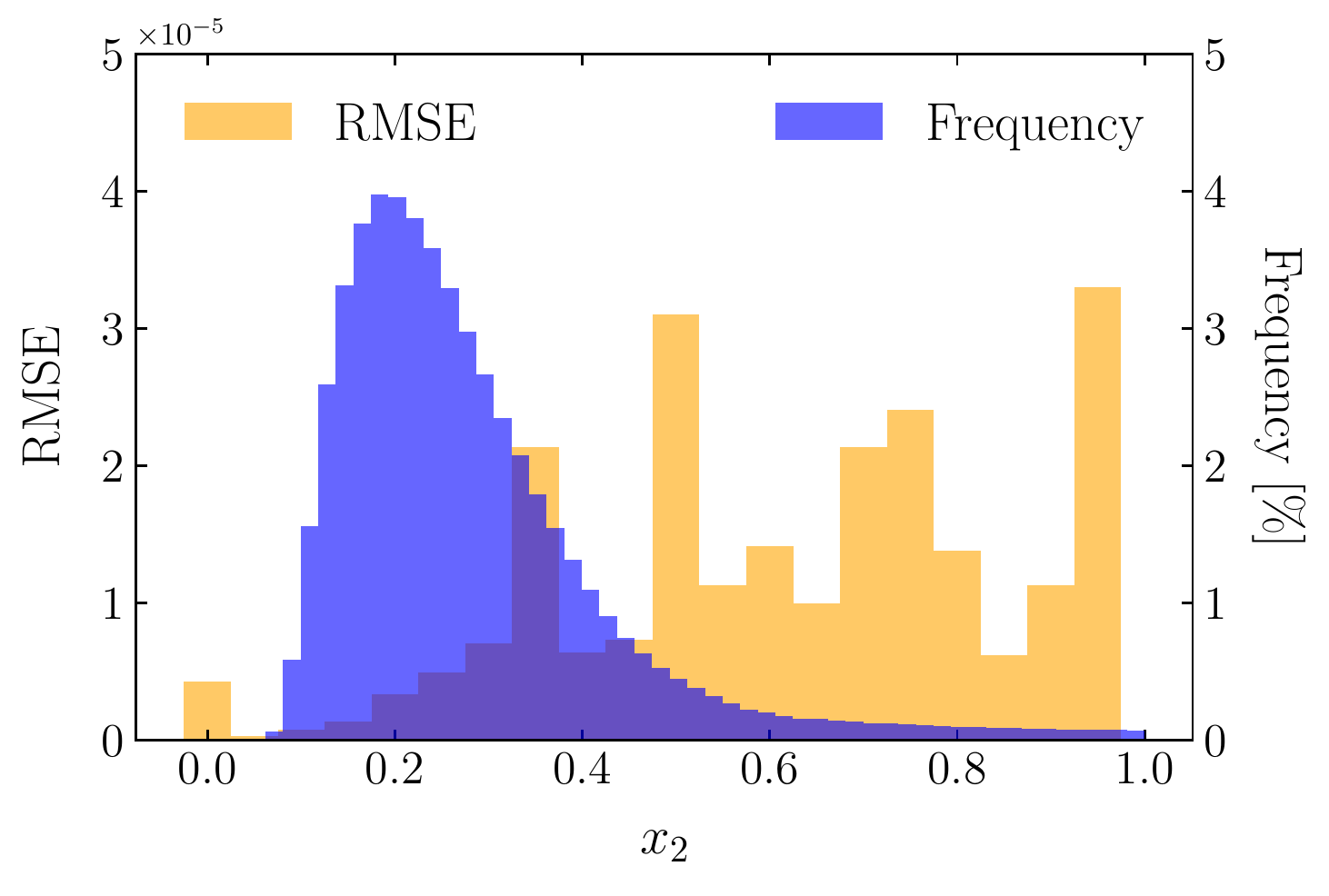}
     \end{subfigure}
     \caption{Root mean squared error (RMSE) of the NN ensemble approach in comparison to \njet~as a function of $x_1$ and $x_2$, and the frequency of points with these values in the training dataset.}
     \label{fig:5_x_1_x_2_error}
\end{figure}

Finally, although the cross section and differential distributions provide a means to test the robustness of our approach against the additional weights introduced during event generation, we can more explicitly single out the effects of the PDF weights by calculating the NN ensemble error as a function of the momentum fractions, $x_1$ and $x_2$, of the initial state partons. \cref{fig:5_x_1_x_2_error} shows the root mean squared error (RMSE) of the NN as a function of these variables, along with the frequency of points as they appear in the training dataset. As expected, the ensemble performs better in locations with more points, and we only see the RMSE grow more significantly in the regions of low-statistics. Since the gluon PDF falls off as $x$ approaches one, and peaks in the low $x$ region, this provides another test of ensemble robustness during the external introduction of PDF weights. 

\subsubsection{Aside: VEGAS grid optimisation}

\begin{figure}
    \centering
    \includegraphics[width=0.5\textwidth]{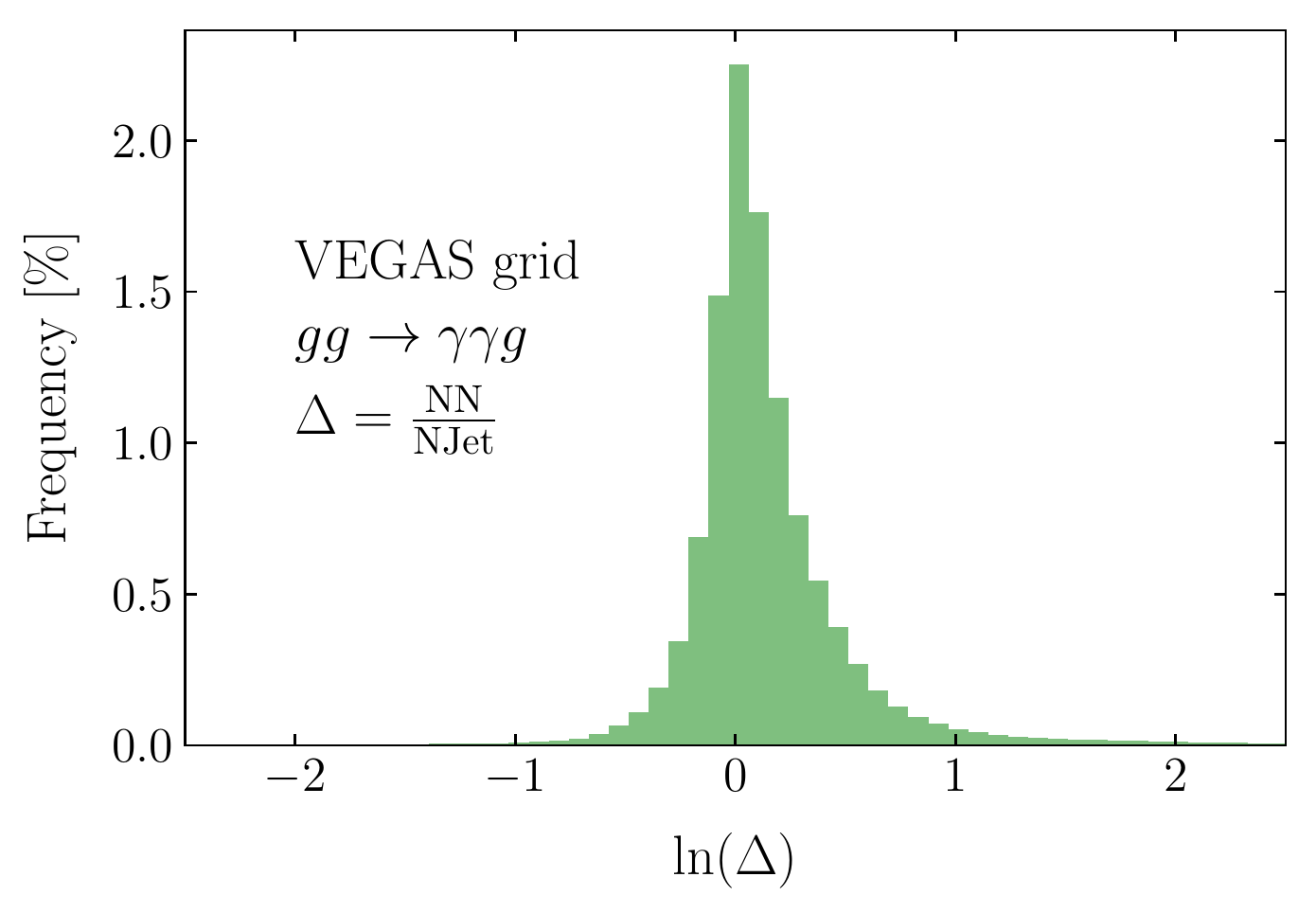}
    \caption{NN/\njet~errors for the $2\rightarrow3$ scattering process using a VEGAS optimised integration grid.}
    \label{fig:5_error_vegas}
\end{figure}

The results presented so far have used a unit integration grid and RAMBO integrator in order to be process agnostic in the phase-space sampling. As mentioned in \cref{sec:integration}, however, it is common to use importance sampling and other optimisation techniques to speed up integration conversion. To test the robustness of our approach to these alternative integrators, we use VEGAS during the optimisation grid generation stage. \cref{fig:5_error_vegas} shows the error plots for the $2\rightarrow3$ scatting process using this optimisation setup while keeping all other parameters setup parameters fixed. Here, we see that the shape exhibited in the error plots is similar to that of the unit grid shown in \cref{fig:5_error_unit}, although slightly broader around the peak. This is likely due to the the larger number of points placed in the divergent regions by the VEGAS integrator. The cross section was also found to be in excellent agreement, with \njet~giving $4.151 \times 10^{-6} \pm 1.1 \times 10^{-8}~\text{pb}$, and the ensemble giving $4.22 \times 10^{-6} \pm 8 \times 10^{-8}~\text{pb}$.

\subsection{\texorpdfstring{$gg \rightarrow \gamma \gamma gg$}{4g2A}}

\begin{figure}
    \centering
    \includegraphics[width=0.5\textwidth]{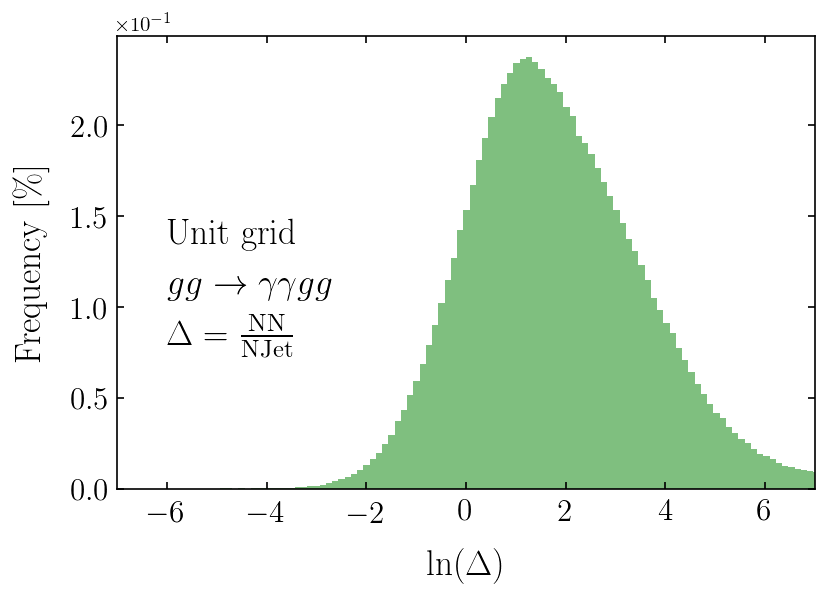}
    \caption{NN/\njet~errors for the $2\rightarrow4$ scattering process using a unit integration grid.}
    \label{fig:6_error_unit}
\end{figure}

We now turn to investigate the $gg \rightarrow \gamma \gamma gg$ process. Analytic expressions for this process are not available and the numerical implementation is significantly more computationally expensive than for the equivalent $2\rightarrow3$ process (see \cref{sec:timing}). Integration grid optimisation is therefore highly inefficient, and so for the remainder of this section a unit grid will be used. To test generalisability, the NN setup is as in \cref{sec:results_5}, with the only change being in the chosen value of $y_{p} = 0.001$. At higher multiplicity, a greater proportion of points fall within the divergent region, $\mathcal{R}_{\text{div}}$, however, this can hinder model performance by unbalancing the training regime. It is therefore reasonable to aim to keep the proportion of points in this region approximately constant throughout our experiments which is achieved by lowering the value of $y_{p}$ (see \cref{app:delta_tuning} for more details).

\cref{fig:6_error_unit} shows the performance of our trained NN ensemble at the matrix element level. As expected, the performance has decreased relative to the $2\rightarrow3$ process shown in \cref{fig:5_error_unit}, yet the error distribution is still found to be approximately Gaussian, although with a shifted mean. Despite this, the cross section calculated using the NN ensemble --- $4.5 \times 10^{-6} \pm 6 \times 10^{-7}$ pb --- is found to be in excellent agreement with that derived from \njet~--- $4.9 \times 10^{-6} \pm 5 \times 10^{-7}$ pb. 
This suggests that although there are several points where the ensemble approach performs poorly, particularly in comparison to the $2\rightarrow3$ process, these are largely in the divergent region and found to not affect the cross-section calculation too greatly.

\begin{figure}
     \centering
     \begin{subfigure}[b]{0.42\textwidth}
         \centering
         \includegraphics[width=\textwidth]{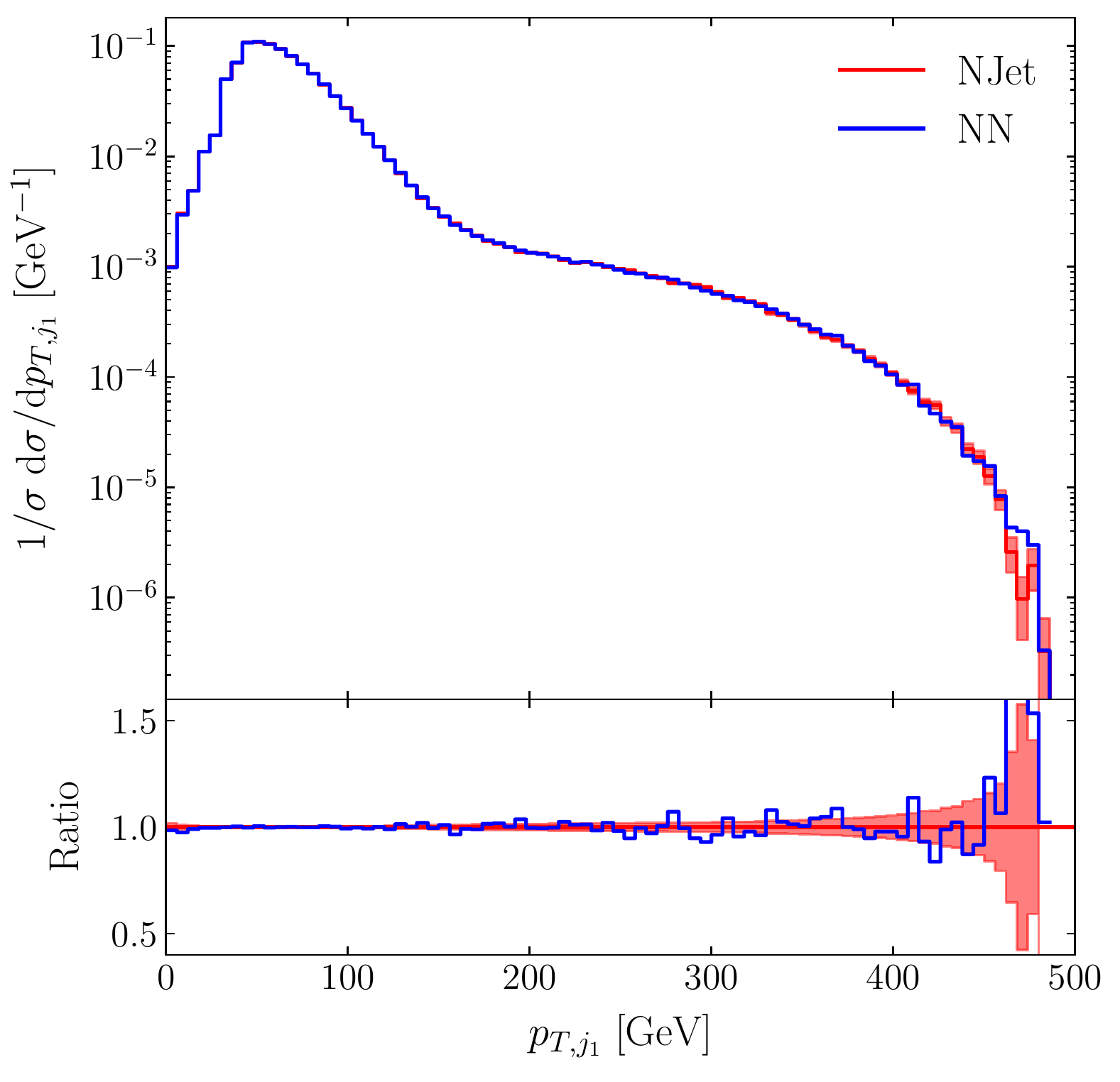}
     \end{subfigure}
     \hfill
     \begin{subfigure}[b]{0.42\textwidth}
         \centering
         \includegraphics[width=\textwidth]{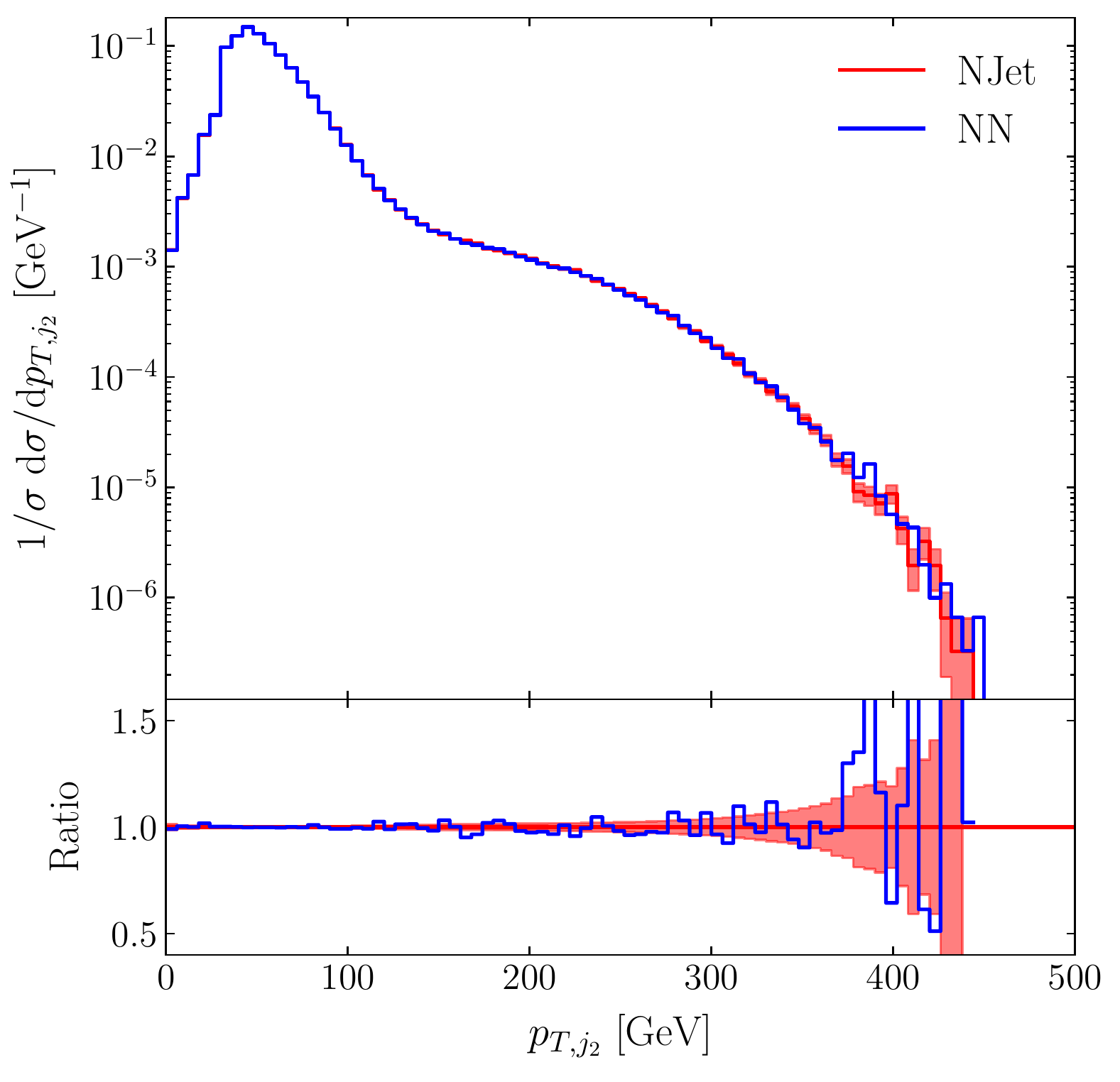}
     \end{subfigure}
     \vfill
     \begin{subfigure}[b]{0.42\textwidth}
         \centering
         \includegraphics[width=\textwidth]{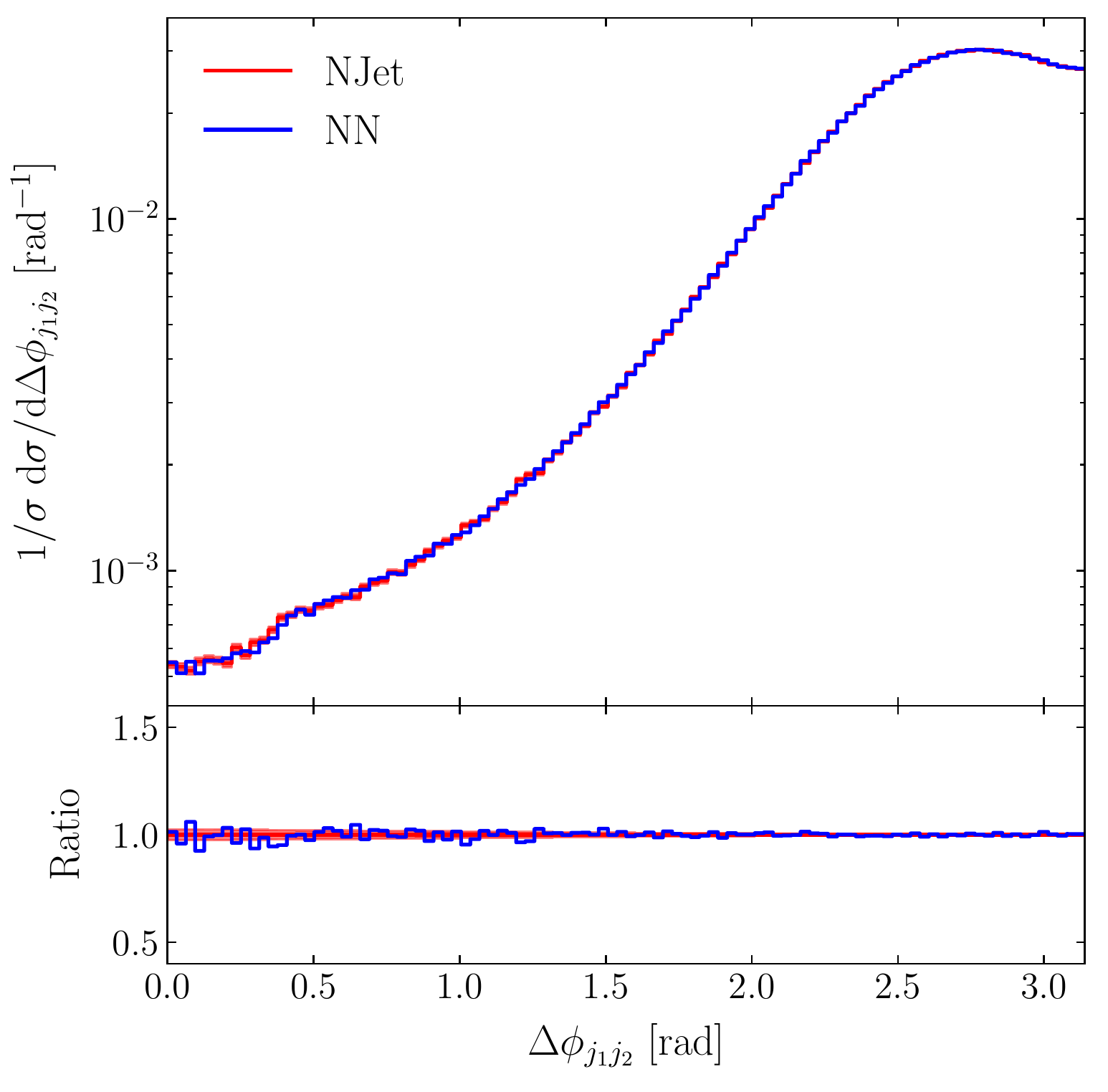}
     \end{subfigure}
     \hfill
     \begin{subfigure}[b]{0.42\textwidth}
         \centering
         \includegraphics[width=\textwidth]{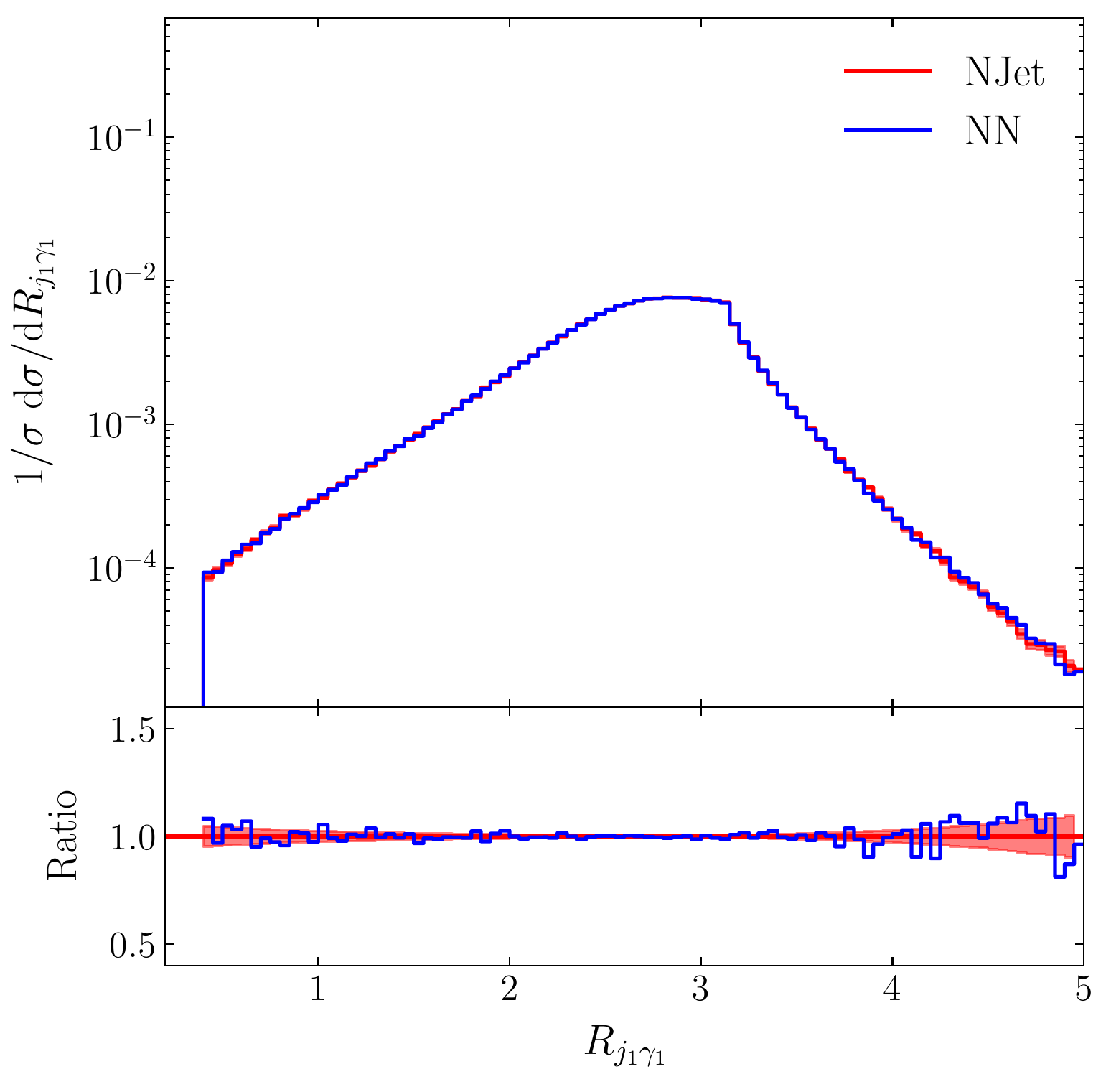}
     \end{subfigure}
     \vfill
     \begin{subfigure}[b]{0.42\textwidth}
         \centering
         \includegraphics[width=\textwidth]{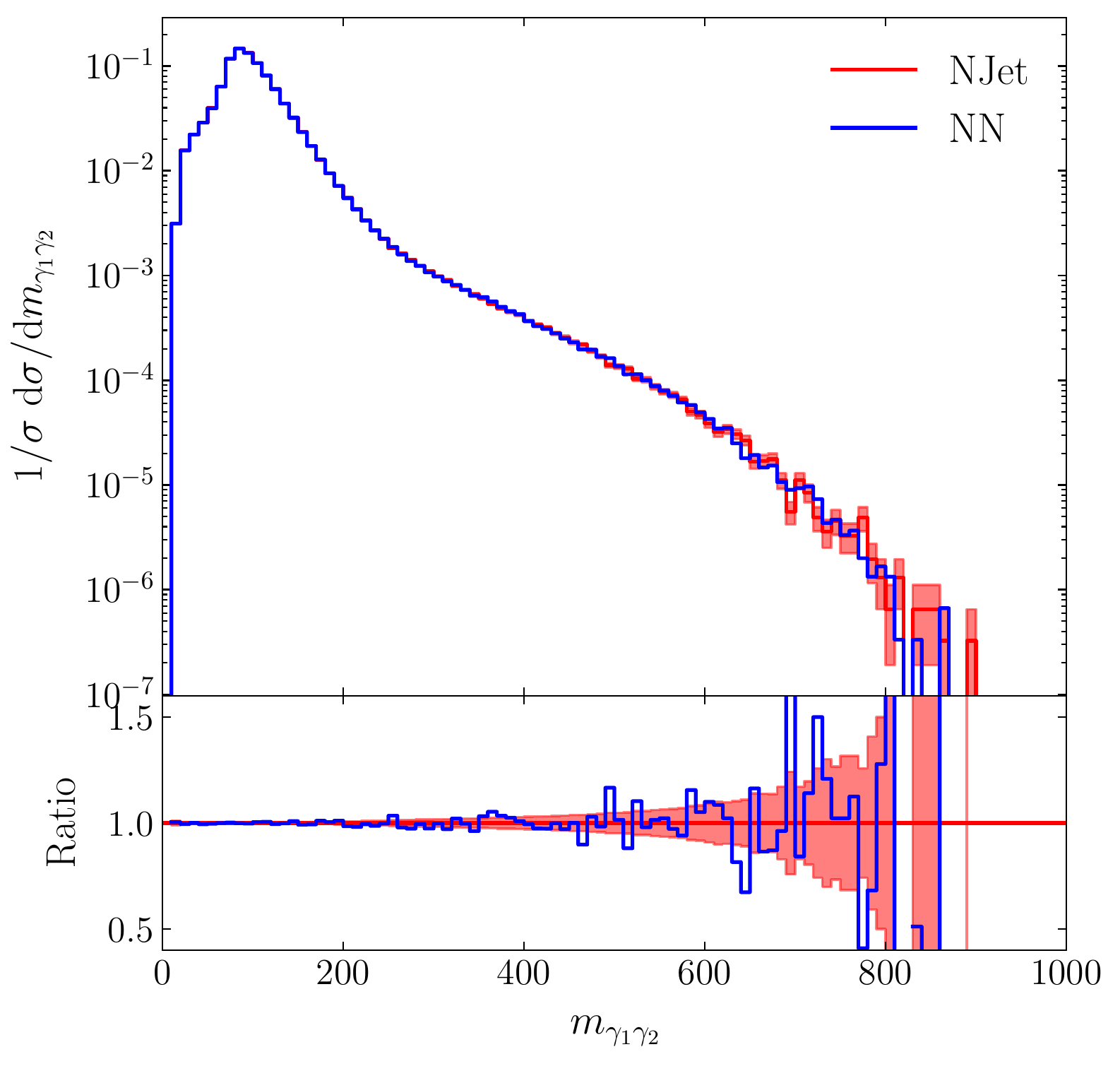}
     \end{subfigure}
     \hfill
     \begin{subfigure}[b]{0.42\textwidth}
         \centering
         \includegraphics[width=\textwidth]{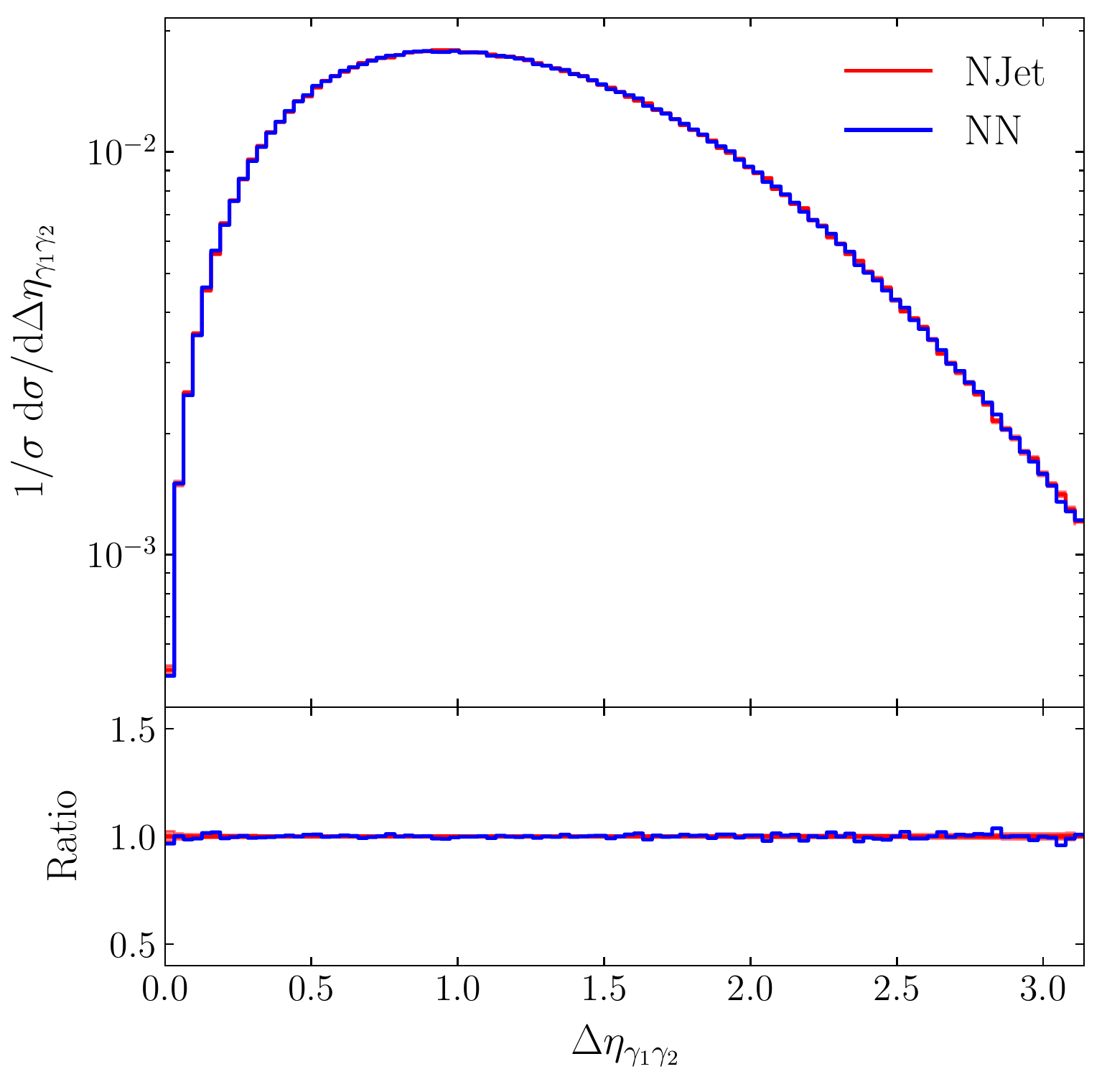}
     \end{subfigure}
     \caption{Differential distributions normalised to the cross section for the $2\rightarrow4$ process comparing \njet~(red) with the NN ensemble (blue). The \njet~results are quoted with MC errors, and the NN results with precision/optimality uncertainties calculated as described in \incite{Badger:2020uow} but which are negligible in comparison.}
     \label{fig:6_differentials}
\end{figure}

\cref{fig:6_differentials} shows the performance of the ensemble approach in six differential slices of phase space. As in the previous example, the ensemble is found to perform well relative to \njet: while noise in the tails of the distributions is still observed, these appear to be reduced in comparison to the $2\rightarrow3$ process. This further supports the assertion that the points where the ensemble performs poorly are suppressed.

\begin{figure}
    \centering
    \includegraphics[width=0.6\textwidth]{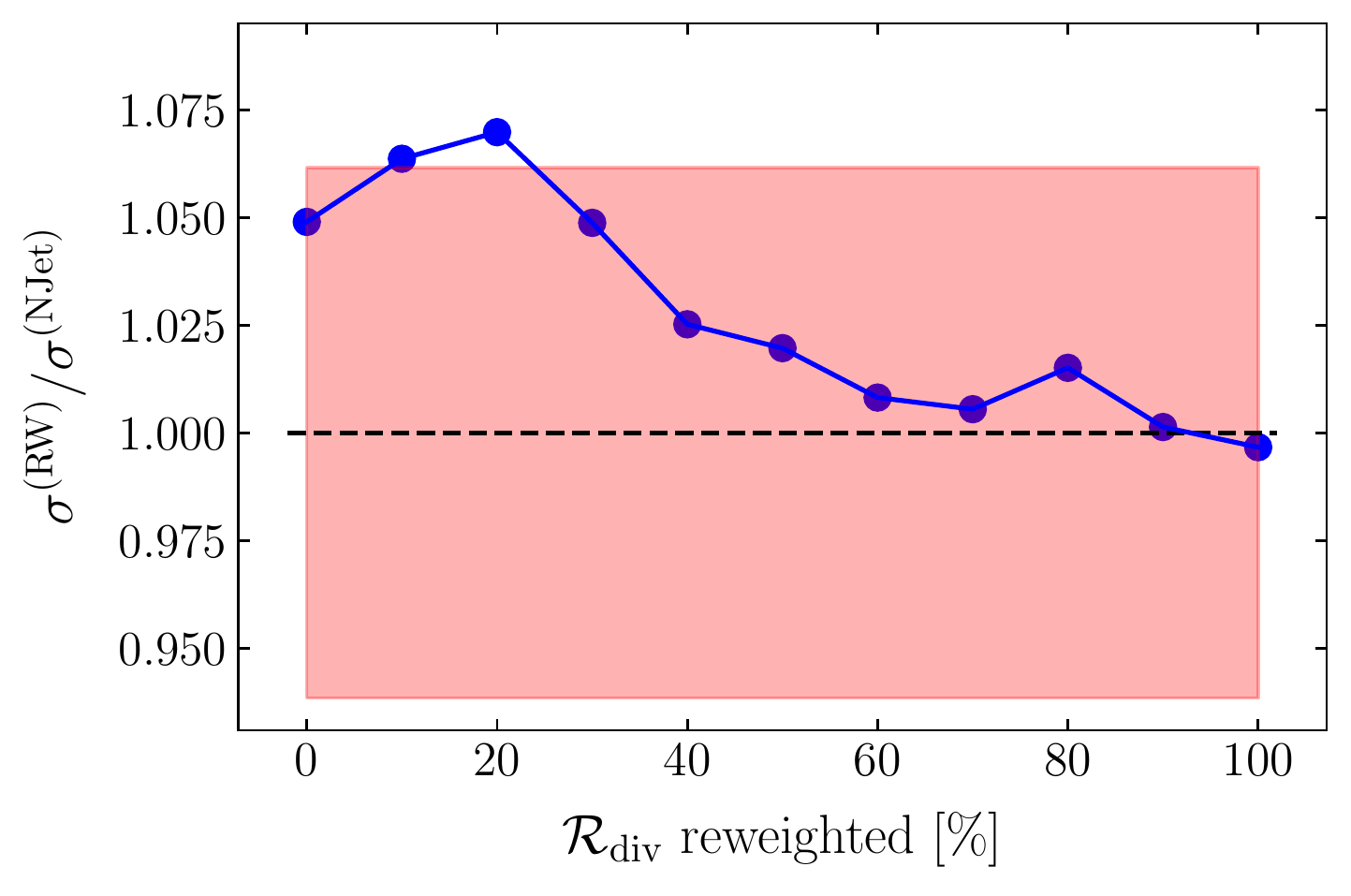}
    \caption{Effect of reweighting points in the divergent region of phase space, $\mathcal{R}_{\text{div}}$, on the ratio between the reweighted cross section, $\sigma^{\text{(RW)}}$, and the cross section calculated using \njet~$\sigma^{\text{(\njet)}}$ for the $2\rightarrow4$ process. In this case, the divergent region comprises approximately 2--3\% of the total phase space (see \cref{app:delta_tuning} for details). The red band shows the MC error on the \njet~result.}
    \label{fig:6_reweighting}
\end{figure}

Given the difference in cross-section values calculated using \njet~and the ensemble approach, we perform reweighting in the divergent region as discussed in \cref{sec:reweighting} and \cref{sec:results_5}. As shown in \cref{fig:6_reweighting}, reweighting in this region can bring the NN ensemble derived cross section closer to the value calculated using \njet. In the case of the $2\rightarrow4$ process, the MC error on the \njet~result is significantly larger for the same number of points compared to the $2\rightarrow3$ process. Given these larger error, and that the ratio $\sigma^{\text{(RW)}}/\sigma^{\text{(\njet)}}$ resides within these errors, it is predictably noisy, yet still converges showing that this approach to reweighting can be generalised across multiple processes.

\subsection{Timing}\label{sec:timing}

We repeat the performance evaluation of \cref{fig:timing-basic} with methods involving error estimation as these are likely to be employed in real-world usage.
For conventional techniques, the dimension scaling test is a standard way to estimate error on the result and introduces a second matrix element call for each phase-space point evaluation.
As discussed in \cref{sec:results_5}, we propose running 20 NN ensembles for each point to obtain a mean with standard error.

\begin{figure}
    \centering
    \includegraphics[width=0.6\textwidth]{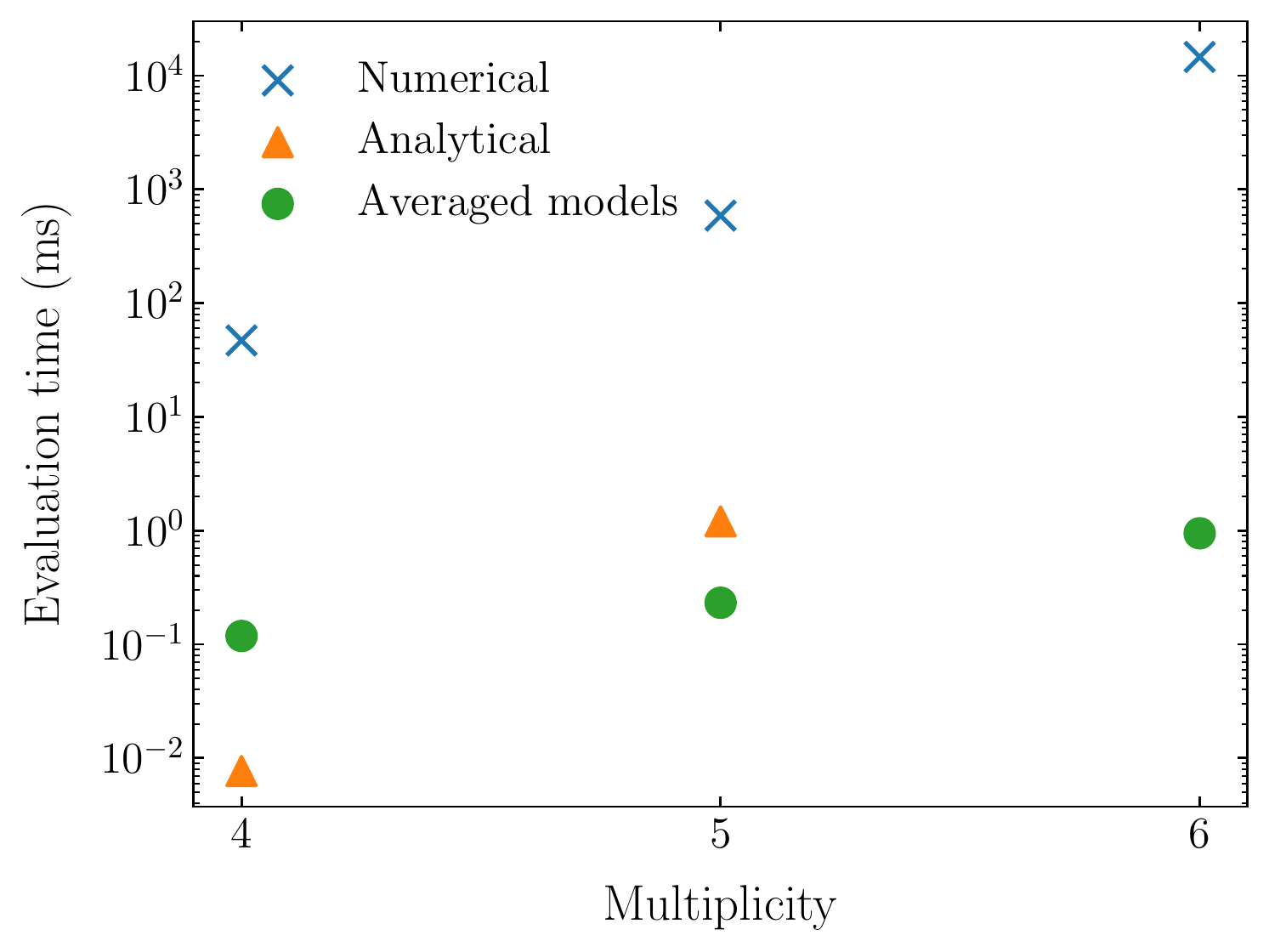}
    \caption{
    Typical per-point call times for the set of NN ensembles and scaling tests with numerical and analytical techniques against the number of legs.
    Compared to \cref{fig:timing-basic}, this incurs a twofold cost on the conventional methods and multiplies the single NN ensemble time by 20.
    Analytical methods are fastest at $2\to2$ and NNs do not offer a dramatic improvement at $2\to3$ either, but their fast call time and weak dependence on the number of variables (which scales with multiplicity) win out at high multiplicity.
    At $2\to4$, where no analytical expression is available and extrapolation suggests it would be comparable in call time to numerics, our ML approach is four orders of magnitude faster than the numeric call.
    }
    \label{fig:timing-detail}
\end{figure}

The results, shown in \cref{fig:timing-detail}, demonstrate the per-point speedup in using amplitude NNs in practice. For the $2\to4$ process, where amplitude calls dominate conventional simulation time, a $10^4$ times speedup in amplitude calls is observed which renders the inference stage as negligible in the total time of our NN-based simulation pipeline. Indeed, in comparison to the numerical calculation of the matrix element, the training time of the NN ensemble can also be considered negligible, meaning the total speed up in the overall simulation time is of the order $N_{\text{infer}}/N_{\text{train}}$ --- the ratio of the number of inference point to the number of points in the training dataset.

%% file: conclusions.tex
\section{Conclusions \label{sec:conclusions}}

In this article we provided further evidence that NNs provide a
general framework for the optimisation of high multiplicity observables at
hadron colliders. We extended preliminary studies~\cite{Badger:2020uow}
for electron-position scattering to hadron-hadron collisions and provided a
general interface to the \sherpa~MC event generator for NNs trained with
the \njet~amplitude library. We focused on the loop-induced processes
$gg\to\gamma\gamma+n(g)$ which cause problems for conventional phase-space generation methods
and require the computation of expensive scattering amplitudes.

We trained an ensemble of NNs which divide the scattering
amplitudes into IR divergent sectors according to the FKS mapping and
find excellent agreement between distributions generated with the networks and
those generated with the conventional approach. Errors from the NN
were included through variations of training parameters and show good agreement
with the direct comparisons. We also showed that by reweighting the generated
events according to their divergence structure, the accuracy of the simulation
could be improved at a rather low computational cost. This step also provided
additional confidence in the inference of the network.

We saw, especially for the $2\to 4$ process, a good improvement in the total
simulation time. Since the calls to the scattering amplitudes dominated the
total time, the speedup was given by the ratio of the number of points used in training to the total
number of calls used in the full simulation (during event generation). This came out around a factor of 30. While this was good
to see, it is not the limit of the optimisation. If the trained networks can be
used for many subsequent simulations with different kinematic cuts, the overall
improvement would be much greater. We showed that our networks
reproduce distributions with different cuts in the transverse momentum without the
requirement for retraining, which was very encouraging. Additional improvements to the inference stage (event
generation with the network) could be provided through GPUs which could be
important if a large number of variations in the cut analysis were required. The distribution of the trained
networks was also simpler than the large quantity of data generated with Root
Ntuples~\cite{Bern:2013zja,Heinrich:2016jad}, another technique for
optimising the information that can be extracted from expensive simulations.

There remain open questions of course. It would be very interesting to apply the
technique to the more intricate problem of real radiation event generation since NLO
and NNLO simulations are often dominated by these contributions. It may also be beneficial to make connections between the amplitude-level approach taken in
this article and those focusing on phase space or complete simulation
including parton showering and even detector simulation.

We hope that these studies will help to develop a general framework that can be used in future experimental analysis.

%% file: appendix.tex
\begin{appendix}
\section{Hyperparameter tuning}\label{app:hyperparameter}

Hyperparameter tuning was performed on a dataset of 1M points (derived independently from the datasets used for validation and testing in \cref{sec:results}) to explore optimal data processing and model parameter choices. Given the computational expense of generating data, this was only done for the $2 \rightarrow 3$ process.

We tested different model architecture constructions (changing the number of hidden layers and/or the number of nodes in each hidden layer), data preprocessing methods, and model loss functions. All other training parameters are as described in \cref{sec:nn_setup}. For data preprocessing methods, we tested input variable standardisation, i.e.~the training and validation data input variables are each standardised to have zero mean and unit variance, and normalisation, i.e.~the training and validation data input variables are each normalised according to min/max normalisation

\begin{equation}
    x^* = \frac{x - \text{min}(X_{\text{train}})}{\text{max}(X_{\text{train}}) - \text{min}(X_{\text{train}})}
\end{equation}

\noindent where $x \in X_{\text{train}}$, $X_{\text{train}} \subset \mathbb{R}$ is the set of training data for a given input variable, and $x^*$ is the input variable normalised from $x$. This procedure means the dataset is normalised such that $x^* \in [0,1]$ and therefore encourages a positive-definite output. When using the standardisation preprocessing step, we use hyperbolic-tangent activation functions in the hidden layers, but for normalisation we use rectified linear units (ReLU) \cite{nair2010recified}. This latter choice is to further encourage a positive-definite output and also aims to increase the rate of convergence.

It should be noted that a clear limitation of the positive-definite conditioning of the normalisation procedure is a reliance on the following conditions:

\begin{align}
    \text{min}(X_{\text{train}}) &= \text{min}(X_{\text{train}} \cup X_{\text{test}}),\\
    \text{max}(X_{\text{train}}) &= \text{max}(X_{\text{train}} \cup X_{\text{test}}),
\end{align}

\noindent where $X_{\text{test}} \subset \mathbb{R}$ is the set variable inputs derived from the testing data, and therefore $X_{\text{train}} \cup X_{\text{test}}$ represents the combination of the training, validation and testing sets. Since the performance gain from using the ML approach is that the training and validation sets combined are much smaller than the testing set, the above conditions are likely to break down as $n(X_{\text{train}}) \gg n(X_{\text{test}})$.

Two model loss functions were tested during hyperparameter tuning. The first was the mean squared error (MSE)

\begin{equation}
    L = \frac{1}{n}\sum^{n}_{i=1}(f(x_{i}) - y_i)^2
\end{equation}

\noindent where $n$ is the number of training points, $f: \mathbb{R}^d\to\mathbb{R}$ is the function describing the neural network, $x_i$ is the $i^\text{th}$ $d$-dimensional input data (here $d = 4n$), and $y_i$ the corresponding target variable.
The second is the mean squared logarithmic error (MSLE)

\begin{equation}
    L = \frac{1}{n}\sum^{n}_{i=1}(\text{log}(f(x_{i})+1) - \text{log}(y_i+1))^2.
\end{equation}

Given the problem of approximating matrix element values for complex scattering processes, the target variable can take on a wide range of values spanning several orders of magnitude. These large values can sometimes be especially important to the cross section and so the MSE's penalisation of large outlier values can be beneficial; however, this might also make the training unstable. We included the MSLE during training to test if reducing the sensitivity to large scale variations in the target value is beneficial.

\begin{table}
\centering
\begin{tabular}{ |c|c|c|c|c|c|c|c|c|c|c|c|c| } 
 \hline
      & \multicolumn{2}{|c|}{Processing} & \multicolumn{4}{|c|}{Layers} & \multicolumn{2}{|c|}{Loss} & \multicolumn{2}{|c|}{Error} \\
 \cline{1-11}
      & Std & Norm          & \makecell{20-40-\\20} & \makecell{30-60-\\30} & \makecell{20-30-\\40-30-\\20} & \makecell{30-40-\\50-40-\\30} & MSE & MSLE & RMSE & RMSLE \\
 \hline
    1 & x &  & x &  &  &  & x &  & $2.828 \times 10^{-5}$ & $2.790 \times 10^{-5}$ \\
    2 & x &  & x &  &  &  &  & x & $3.320 \times 10^{-5}$ & $3.288 \times 10^{-5}$ \\
    3 & x &  &  & x &  &  & x &  & $2.829 \times 10^{-5}$ & $2.790 \times 10^{-5}$ \\
    4 & x &  &  & x &  &  &  & x & $3.820 \times 10^{-5}$ & $3.791 \times 10^{-5}$ \\
    5 & x &  &  &  & x &  & x &  & $2.829 \times 10^{-5}$ & $2.791 \times 10^{-5}$ \\
    6 & x &  &  &  & x &  &  & x & $3.147 \times 10^{-5}$ & $3.113 \times 10^{-5}$ \\
    7 & x &  &  &  &  & x & x &  & $2.830 \times 10^{-5}$ & $2.792 \times 10^{-5}$ \\
    8 & x &  &  &  &  & x &  & x & $3.454 \times 10^{-5}$ & $3.422 \times 10^{-5}$ \\
    9 &  & x & x &  &  &  & x &  &  $2.835 \times 10^{-5}$ & $2.797 \times 10^{-5}$ \\
    10 &  & x & x &  &  &  &  & x & $4.799 \times 10^{-4}$ & $4.802 \times 10^{-4}$ \\
    11 &  & x &  & x &  &  & x &  & $2.835 \times 10^{-5}$ & $2.797 \times 10^{-5}$ \\
    12 &  & x &  & x &  &  &  & x & $7.396 \times 10^{-4}$ & $7.405 \times 10^{-4}$ \\
    13 &  & x &  &  & x &  & x &  & $2.836 \times 10^{-5}$ & $2.797 \times 10^{-5}$ \\
    14 &  & x &  &  & x &  &  & x & $3.414 \times 10^{-4}$ & $3.416 \times 10^{-4}$ \\
    15 &  & x &  &  &  & x & x &  & $2.836 \times 10^{-5}$ & $2.797 \times 10^{-5}$ \\
    16 &  & x &  &  &  & x &  & x & $5.599 \times 10^{-4}$ & $5.604 \times 10^{-4}$ \\
 \hline
\end{tabular}
\caption{Hyperparameter tuning results. Tuning was performed on a fixed training dataset size of 100k points sampled using the RAMBO integrator \cite{Kleiss:1985gy} on a unit integration grid. Performance was measured with respect to both the Root Mean Squared Error (RMSE) and Root Mean Squared Logarithmic Error (RMSLE) so as to avoid biasing the error measure to the optimisation criterion (loss function) chosen.}
\label{tab:hyperparameter_tuning}
\end{table}

The results of the hyperparameter tuning can be found in \cref{tab:hyperparameter_tuning}. Here we see that using data standardisation with an MSE loss function generally produces better results, although there does not seem to be a clear dependence on the model architecture. Given these findings, we choose to train our models using data standardisation with hyperbolic-tangent activation functions, a MSE loss function, and an architecture of 20-40-20. This is consistent with that presented in \incite{Badger:2020uow}.

\section{\texorpdfstring{$y_{p}$}{y p} tuning}
\label{app:delta_tuning}

\begin{figure}[htp]
     \centering
     \includegraphics[width=0.7\textwidth]{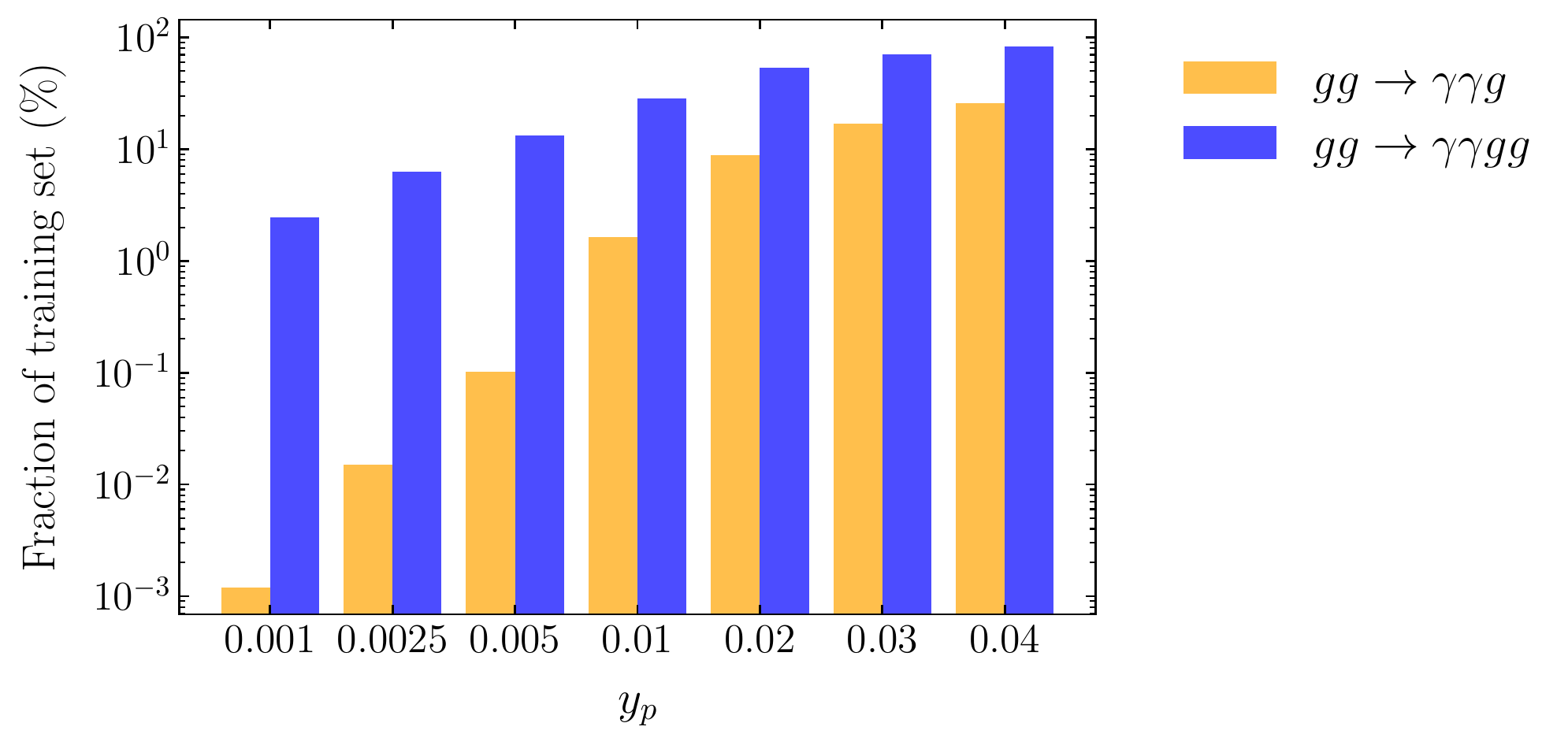}
     \caption{Proportion of the training dataset in the divergent region, $\mathcal{R}_{\text{div}}$, as a function of $y_{p}$ for the $2\rightarrow3$ and $2\rightarrow4$ process.}
     \label{fig:delta_tuning}
\end{figure}

The choice of $y_{p}$ defines the partition between the divergent region of phase space, $\mathcal{R}_{\text{div}}$, and the non-divergent region, $\mathcal{R}_{\text{non-div}}$ (see \cref{eqn:R_div,eqn:R_non_div}). While it may be assumed that having more points in each region is helpful since it provides more data for the networks trained in each region, this is not always the case. Including a mixture of points in the training dataset, with large imbalances in the distribution of different scales, can make the network optimisation procedure increasingly noisy. For this reason, we seek to choose a value of $y_{p}$ which provides a balance between having enough divergent points to learn those regions well, whilst not providing too many points not in this limit and which share similar scales to points in the non-divergent regions of phase space.

To be consistent with prior work \cite{Badger:2020uow}, we initially chose $y_{p} = 0.02$ although the number of points falling into $\mathcal{R}_{\text{div}}$ depends on the multiplicity of the process. As presented in \cref{sec:results_5}, this value was shown to perform well,\footnote{The value of $y_{p} = 0.01$ was also tested and found to be in similarly good agreement.} yet the same value would place a significantly greater proportion of points into the divergent region when another external leg is added (see \cref{fig:delta_tuning}). Instead of choosing the same value of $y_{p}$ for all processes, we aim to select a value which keeps the proportion of points in the divergent region at the level of 2--8\% of the whole phase space sampled. We choose a value of $y_{p} = 0.02$ for the $2\rightarrow3$ process, and $y_{p} = 0.001$ for the $2\rightarrow4$ process.\footnote{A value of $y_{p} = 0.0025$ for the $2\rightarrow4$ process would also allow for this; however, at high multiplicity the lower value of this cut provided more optimal performance.}

\section{Comparison with the naive setup}
\label{app:single_comparison}

\begin{figure}[htp]
     \centering
     \begin{subfigure}[b]{0.49\textwidth}
         \centering
         \includegraphics[width=\textwidth]{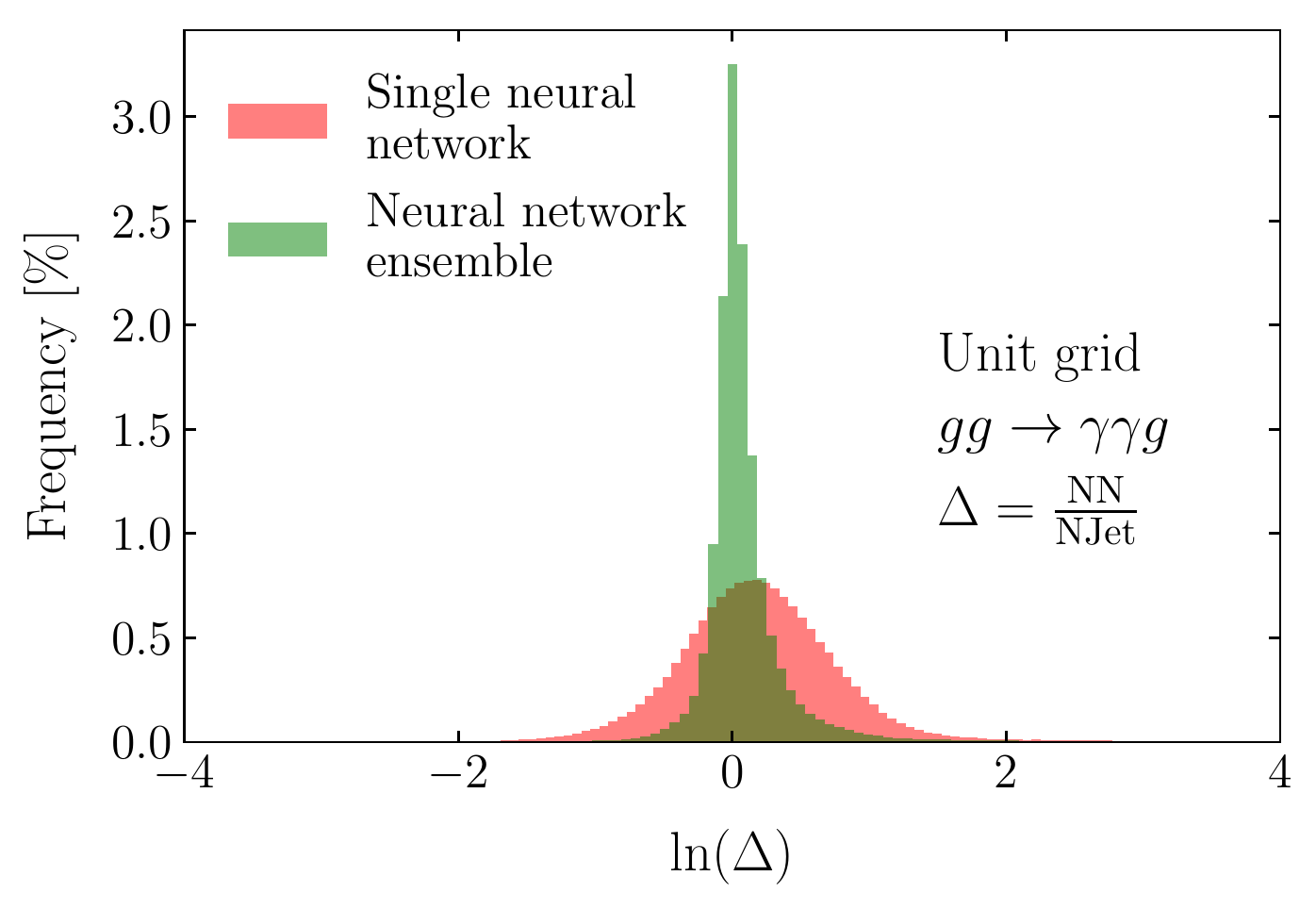}
         \caption{Unit integration grid.}
         \label{fig:5_error_unit_joint}
     \end{subfigure}
     \hfill
     \begin{subfigure}[b]{0.49\textwidth}
         \centering
         \includegraphics[width=\textwidth]{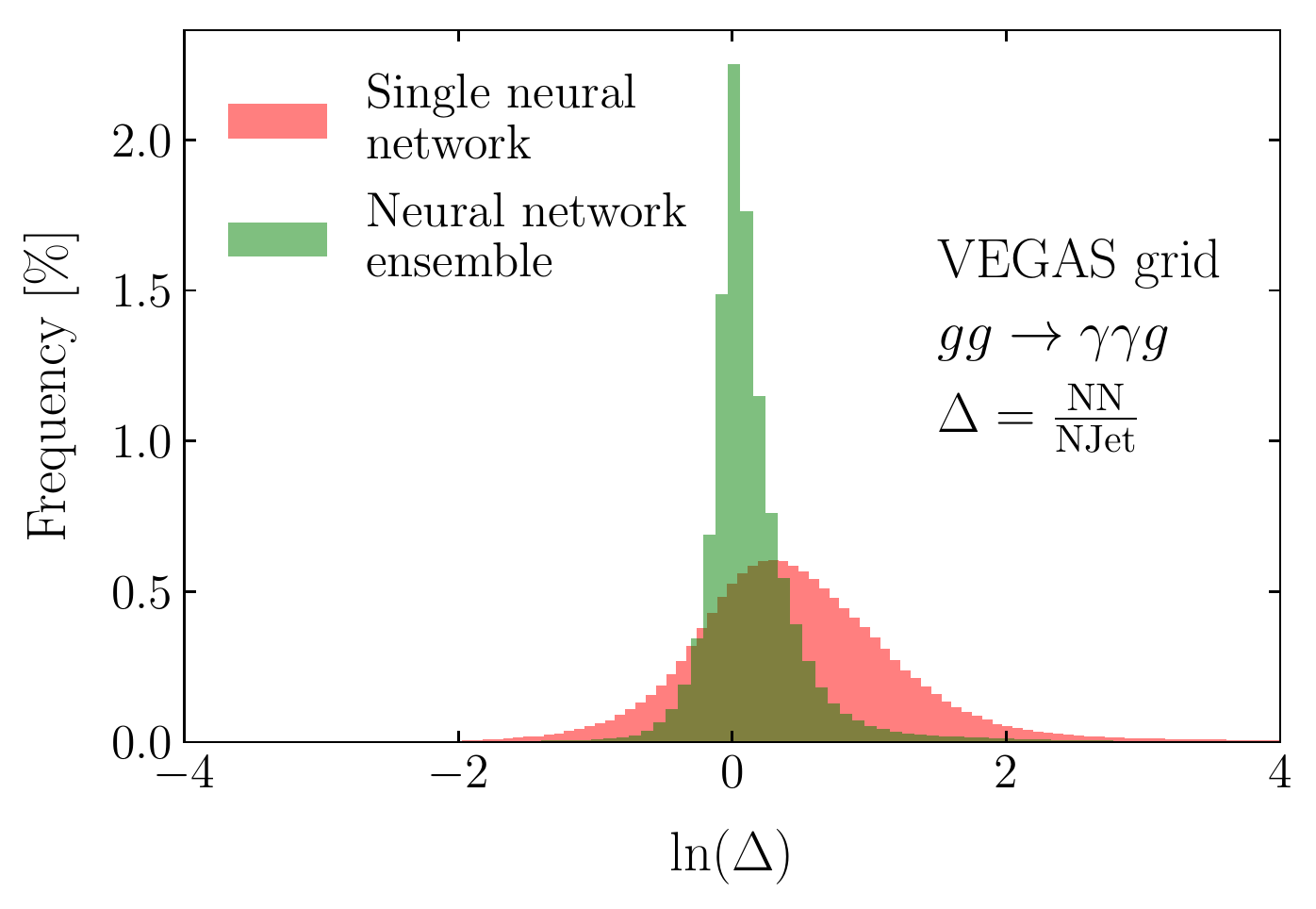}
         \caption{VEGAS integration grid.}
         \label{fig:5_error_vegas_joint}
     \end{subfigure}
     \caption{Comparison of NN/\njet~errors between the single NN and NN ensemble approaches for the $2\rightarrow3$ scattering process using different integration grids.}
     \label{fig:5_error_unit_vegas_joint}
\end{figure}

Throughout this paper, all results presented using a ML approach have used the NN ensemble methodology. This approach had been shown previously to outperform a naive single NN trained over the whole of phase space for $e^+e^-$ collisions \cite{Badger:2020uow}. In particular, the motivation for this approach was enhanced performance in handling real emission, IR singular regions of phase space, which similarly occur in the processes studied in this work, especially at high multiplicity. For completeness, we perform a similar comparison on the $2\rightarrow3$ gluon-initiated diphoton processes; we do not compare on the $2\rightarrow4$ process as it is computationally expensive to do so and it is a natural higher multiplicity extension of the $2\rightarrow3$ process.

\cref{fig:5_error_unit_vegas_joint} shows the matrix level error analysis of the $2\rightarrow3$ scattering process using both a unit and VEGAS optimisation grid. In both cases, the error distribution for the single NN approach has a significantly broader character than the ensemble method. This demonstrates that the findings of described in \incite{Badger:2020uow} are consistent with those presented in this study.

\end{appendix}